\documentclass[aps,twocolumn,superscriptaddress]{revtex4-2}

\usepackage{amsthm}
\usepackage{amsmath,bm}
\usepackage{amssymb}
\usepackage{amsfonts}
\usepackage{graphicx}
\usepackage{txfonts}
\usepackage{xcolor}
\usepackage{float}
\usepackage{braket}
\usepackage[colorlinks=true,linkcolor=blue,citecolor=blue,urlcolor=blue]{hyperref}
\usepackage[capitalise]{cleveref}
\usepackage{bbm}
\usepackage{changes}

\newcommand{\ie}{\textit{i.e.} }

\newcommand\LDSeq{\mathrel{\stackrel{\makebox[0pt]{\mbox{\normalfont\tiny LDS}}}{=}}} 

\newcommand{\iu}{{i\mkern1mu}} 

\begin{document}

\title{Nonlocality, Integrability and Quantum Chaos in the Spectrum of Bell Operators}

\author{Albert Aloy}
\email{albert.aloy@oeaw.ac.at}
\affiliation{Institute for Quantum Optics and Quantum Information, Austrian Academy of Sciences, Boltzmanngasse 3, A-1090 Vienna, Austria}
\affiliation{Vienna Center for Quantum Science and Technology (VCQ), Faculty of Physics, University of Vienna, Vienna, Austria}

\author{Guillem M\"uller-Rigat}
\affiliation{ICFO-Institut de Ciencies Fotoniques, The Barcelona Institute of Science and Technology, Castelldefels (Barcelona) 08860, Spain.}

\author{Maciej Lewenstein}
\affiliation{ICFO-Institut de Ciencies Fotoniques, The Barcelona Institute of Science and Technology, Castelldefels (Barcelona) 08860, Spain.}
\affiliation{ICREA, Pg.~Lluís Companys 23, 08010 Barcelona, Spain.}

\author{Jordi Tura}
\email{tura@lorentz.leidenuniv.nl}
\affiliation{$\langle aQa^L \rangle$ Applied Quantum Algorithms, Universiteit Leiden}
\affiliation{Instituut-Lorentz, Universiteit Leiden, P.O. Box 9506, 2300 RA Leiden, The Netherlands}

\author{Matteo Fadel}
\email{fadelm@phys.ethz.ch}
\affiliation{Department of Physics, ETH Z\"{urich}, 8093 Z\"{urich}, Switzerland}

\begin{abstract}
We introduce a permutationally invariant multipartite Bell inequality for many-body three-level systems and use it to investigate a connection between Bell nonlocality and (lack of) quantum chaos. An associated Bell operator is then defined via Born's rule, mapping the conditional probabilities of the Bell inequality to quantum measurement operators. This allows us to interpret the Bell operator as an effective Hamiltonian, which we use to analyze its spectral statistics across different SU(3) irreducible representations and measurement choices. 
Surprisingly, we find that, in every irreducible representation exhibiting nonlocality, the measurement settings yielding maximal violation result in a Bell operator with Poissonian level statistics, thus signaling integrable behavior. 
This integrability is both unique and fragile, since generic or slightly perturbed measurements lead to the Wigner-Dyson statistics associated with chaotic behavior. 
Through further analysis, we are able to identify an emergent parity symmetry in the Bell operator near the point of maximal violation, providing an explanation for the observed regularity in the spectrum. 
These results suggest a deep interplay between optimal quantum measurements, non-local correlations, and integrability, opening new perspectives at the intersection of Bell nonlocality and quantum chaos. 
\end{abstract}

\maketitle

\section{INTRODUCTION}

Quantum nonlocality, most famously manifested through the violation of Bell inequalities \cite{Bell1964}, enables information-processing tasks that cannot be achieved within classical physics \cite{BrunnerRMP2014}. Over the past decades, considerable theoretical and experimental progress has been made in the detection and characterization of nonlocality in many-body systems. 
These advances not only extend experimental frontiers but also open the door to attributing new physical meaning to nonlocal correlations, positioning them as potential indicators of complex behavior in quantum many-body and chaotic systems.
For example, recent developments have unveiled connections between nonlocal correlations and macroscopic properties such as phase transitions \cite{FadelQuantum2018}, quantum criticality \cite{PigaPRL2019}, and metrological advantage \cite{FroewisArxiv2018}. Yet, our understanding of how nonlocal correlations manifest in systems beyond spin-$1/2$ particles, and how they relate to complex physical behavior, remains limited.

A central challenge is the exponential scaling of Bell scenarios with the number of parties, measurement settings, and outcomes \cite{ChazelleDCG1993}. For spin-1/2 ensembles, this barrier has been mitigated by exploiting symmetry and focusing on low-order correlators, leading to experimentally accessible Bell inequalities with permutational invariance \cite{SciencePaper,WagnerPRL2017,BaccariPRA2019,FadelPRL2017,GuoPRL23}. These approaches have enabled detection of Bell correlations in systems containing up to $5\cdot10^5$ particles \cite{SchmiedScience2016,EngelsenPRL2017}. In contrast, analogous methods for higher-spin particles, where each subsystem has three or more outcomes, are far less developed, and no experimental demonstration of Bell correlations in such systems has yet been reported.

\begin{figure}[h!]
\centering
\includegraphics[width=0.9\linewidth]{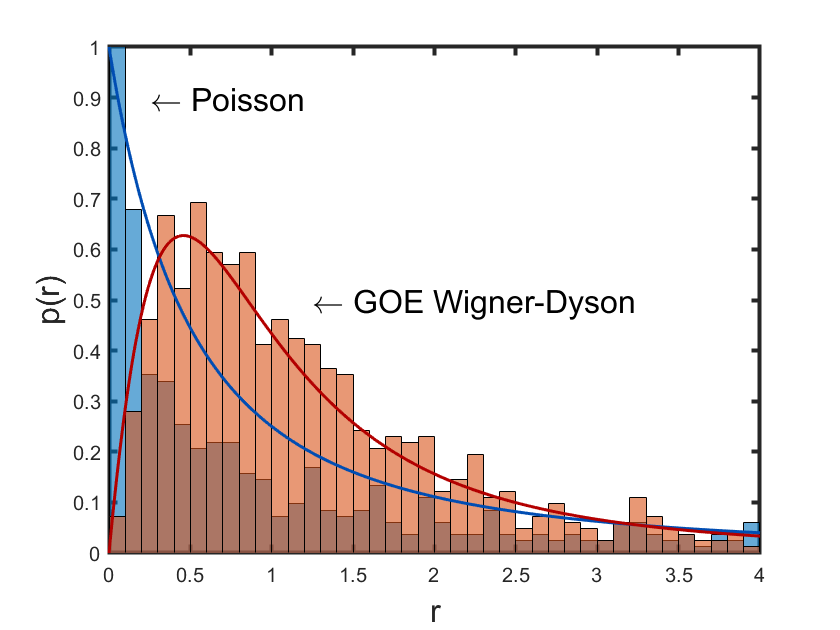}
\caption{Ratio of Consecutive level Spacings (RCS) for the Bell operators associated with the PIBI~(\ref{eq:PIBI}) for $n=25$, obtained with optimal (blue) and random (red) measurement settings. Spectra are shown for the irreducible representation $(p,q)=(21,2)$, chosen for illustration, which has $825$ eigenvalues in the symmetric subspace. Solid curves are fits to the interpolating RCS function in Eq.~\eqref{eq:RCSinterpolation}, which spans Poisson statistics ($\lambda = 0$, indicating integrability) to Wigner-Dyson statistics ($\lambda = 1$, indicating chaos). In the Wigner-Dyson case, the Gaussian Orthogonal Ensemble (GOE) seems to provide a better fit than the Gaussian Unitary Ensemble (GUE), suggesting that time-reversal symmetry is preserved in the chaotic regime.} 
\label{fig1}
\end{figure}

The case of spin-1 ensembles is particularly compelling. Three-level many-body systems naturally arise in ultracold atomic platforms \cite{LawPRL98,HamleyNat12,KitzingerPRA21,OurFriends}, play a central role in the physics of exotic quantum phases \cite{haldane1983continuum,HaldanePRL83,AKLTPRL87}, and feature prominently in collective models of nuclear physics \cite{Elliott1958}. This has spurred efforts to simulate qudit Hamiltonians using trapped ions, superconducting circuits and ultracold atoms. From a theoretical perspective, SU(3) models already host intrinsic quantum chaotic dynamics, in contrast to SU(2) models that typically require external driving \cite{KickedRotator,HaakeKickedTop}. This makes SU(3) systems natural platforms to explore the interplay between dynamical complexity, in particular the integrable-to-chaotic transition, and foundational quantum information features such as nonlocal correlations.

While classical chaos is associated with an exponentially fast growing distance between phase-space trajectories, quantum chaos is often diagnosed through spectral properties of the Hamiltonian. 
According to the Bohigas–Giannoni–Schmit (BGS) conjecture \cite{bohigas1984characterization}, quantum systems with classically chaotic counterparts exhibit energy spectra resembling those of Random Matrix Theory \cite{mehta2004random,Haake2018}. Such insight has led to use spectral statistics to distinguish integrable and chaotic dynamics, with Poisson and Wigner-Dyson distributions serving as key indicators \cite{Haake2018}. This spectral lens provides a powerful diagnostic tool that has rarely been explored in the context of Bell operators and nonlocality.

In this work, we explore the interplay between quantum chaos and nonlocality by introducing a permutationally invariant Bell inequality tailored to multipartite spin-1 systems. We construct the corresponding Bell operator and analyze its spectral properties under different measurement configurations. We find that generic SU(3) representations display spectral signatures of quantum chaos, whereas configurations that maximize the Bell inequality violation exhibit integrable features. Notably, this transition appears to be linked to the emergence of a parity symmetry in the Bell operator near maximal nonlocality detection. These findings reveal a novel connection between maximal violation of Bell inequalities, integrability, and random matrix theory, opening a new direction for understanding nonlocality and quantum correlations in high-dimensional many-body systems.

\section{RESULTS}
\subsection{The Bell inequality and its corresponding Bell operator} 

We begin by introducing the Bell inequality and its associated Bell operator that forms the basis of our analysis. Consider a typical two-setting, three-outcome multipartite Bell scenario \cite{BrunnerRMP2014}: A set of $n$ space-like separated and non-communicating parties labeled by $i\in [n]:=\{1,2,\ldots,n\}$ share an $n$-partite resource (e.g., a quantum state in a Hilbert space composed of $n$ subsystems). Each party $i\in [n]$ performs a local measurement on their subsystem by selecting a measurement setting $x_i\in\{0,1\}$, which specifies the local observable being measured. The resulting outcome $a_i\in\{0,1,2\}$ is recorded, and the process is repeated until sufficient statistics are collected. From the methodology presented in \cite{aloy2024deriving}, one can derive the following three-outcome Permutationally Invariant Bell Inequality (PIBI):
\begin{eqnarray}\label{eq:PIBI}
    && B = (\mathcal{P}_{0|0}+\mathcal{P}_{0|1}+ \mathcal{P}_{1|0}+\mathcal{P}_{1|1}) + \\
    && (\mathcal{P}_{00|00}+\mathcal{P}_{00|11}+\mathcal{P}_{11|00}+\mathcal{P}_{11|11}) - 2(\mathcal{P}_{01|01}+\mathcal{P}_{01|10}) \geq 0  \;,\nonumber
\end{eqnarray}
where $\mathcal{P}_{a|x}=\sum_{i\in [n]}p_i(a|x)$ is the collective one-body conditional probability, with $p_i(a|x)$ denoting the probability that subsystem $i$ yields outcome $a$ given measurement setting $x$. Similarly, $\mathcal{P}_{ab|xy}=\sum_{i\neq j \in [n]}p_{ij}(ab|xy)$ represents the collective two-body conditional probability summing over all possible pairs $i\neq j\in [n]$. In Supplementary Materials~\cref{app:classicalBound} we prove that, under the principles of local-realism \cite{EPR35,Bell1964}, Eq.~\eqref{eq:PIBI} has classical bound $\beta_c = 0$, classifying it as a Bell inequality. Therefore, observing any violation of the Bell inequality ($B<0$) signals non-local correlations (or simply, \textit{nonlocality}).

Quantum theory allows for correlations that go beyond the principles of local-realism. To demonstrate that Bell inequality~\eqref{eq:PIBI} can indeed be violated in quantum mechanics, one must find suitable quantum states and measurements yielding $B<0$. To this end, we associate each measurement setting in~\eqref{eq:PIBI} with its quantum representation as a self-adjoint operator, allowing Eq.~\eqref{eq:PIBI} to be written as the expectation value of a Bell operator $\mathcal{B}$ via Born's rule, \textit{i.e.} $\braket{\mathcal{B}}=\mathrm{Tr}\left[\rho\mathcal{B}\right]=B$ for some global quantum state $\rho$. For example, let $\left\lbrace E_{a|x}\right\rbrace_{a=0}^2$ be the local Positive Operator-Valued Measurements (POVMs) for setting $x$, with $E_{a|x}\succeq 0$ and $\sum_a E_{a|x}=\mathbb{I}$. Then, one has 
\begin{equation}
p_i(a|x)=\mathrm{Tr}\left[ \rho E_{a|x}^{(i)}\right], \quad p_{ij}(ab|xy)=\mathrm{Tr}\left[ \rho E_{a|x}^{(i)}E_{b|y}^{(j)}\right],
\end{equation}
where $E_{a|x}^{(i)}=\mathbb{I}^{\otimes (i-1)}\otimes E_{a|x}\otimes \mathbb{I}^{\otimes (n-i)}$ is the POVM associated to measurement $x$ yielding outcome $a$ when measuring subsystem $i$, while acting trivially on all other subsystems. Similarly, the products $E_{a|x}^{(i)} E_{a|x}^{(j)}$ denote two-body terms acting nontrivially only on subsystems $i$ and $j$. Therefore, the collective conditional probabilities in \eqref{eq:PIBI} are expectations of the following associated Bell operator:
\begin{eqnarray}
\mathcal{B}=&&\sum\limits_{i\in [n]}\sum_{\substack{a \in \{0, 1\} \\ x \in \{0, 1\}}} E^{(i)}_{a|x} + \sum\limits_{i\neq j \in [n]} \sum\limits_{\substack{a \in \{0, 1\} \\ x \in \{0, 1\}}} \left(E_{a|x}^{(i)} E_{a|x}^{(j)}\right) \nonumber \\
&& - 2 \sum\limits_{i\neq j \in [n]} \left( E_{0|0}^{(i)}  E_{1|1}^{(j)} + E_{0|1}^{(i)} E_{1|0}^{(j)}\right).
\label{eq:bell_operator}
\end{eqnarray} 
where $a=2$ does not explicitly appear due to no-signalling constraints (see Supplementary Methods~\cref{app:classicalBound}) but is present through $E_{2|x}=\mathbb{I}-E_{0|x}-E_{1|x}$.

The Bell inequality~(\ref{eq:PIBI}) can therefore be rewritten as the expectation value of the Hermitian operator $\mathcal{B}$ in~\eqref{eq:bell_operator}. This operator acts on the same $3^n$-dimensional Hilbert space of $n$ qutrits and is constructed solely from sums and products of local measurement operators. Consequently, $\mathcal{B}$ inherits the tensor-product structure and permutation invariance of the underlying Bell scenario. This structural correspondence is what allows us to view $\mathcal{B}$, when expressed in an appropriate symmetry-adapted basis, as an effective many-body Hamiltonian whose blocks correspond to the irreducible symmetry sectors analyzed in the coming sections. See~\cite{TuraPRX2017,FadelQuantum2018} for detailed discussions of many-body Bell operators viewed as effective spin Hamiltonians.


In practice, we parametrize this Bell operator as $\mathcal{B}(\boldsymbol{\theta})$, where $\boldsymbol{\theta}$ is a vector specifying the measurement settings (\textit{e.g.,} spin measurements along two possible directions given by $\boldsymbol{\theta}$ and labelled by $x\in\{0,1\}$).
Under quantum theory, Eq.~\eqref{eq:PIBI} can thus be evaluated as $B=\mathrm{Tr}(\mathcal{B}(\bm{\theta})\rho)$, where $\rho$ is the shared quantum state upon which the measurements specified by $\bm{\theta}$ are performed.

The maximum quantum violation of the inequality is directly related to the minimal eigenvalue $e_\text{min}$ of the associated Bell operator. Specifically, solving the optimization problem $\min_{\boldsymbol{\theta}}e_\text{min}(\mathcal{B}\left(\boldsymbol{\theta})\right)$ gives the minimal quantum value achievable by $B$. A negative minimum corresponds to the maximal quantum violation.
However, solving this minimization over the measurement parameters $\boldsymbol{\theta}$ is far from trivial. To proceed, we require an efficient way to parametrize local projective measurements while respecting the structure of the Bell scenario. Namely that each party chooses between two possible local measurements, each with the same possible three outcomes.

\subsection{Measurements parametrization for optimization}

In higher-dimensional systems such as qutrits, smoothly parametrizing local projective measurements compatible with Eq.~\eqref{eq:bell_operator} is considerably more involved than in the qubit case. 
To address this, we adopt a unitary-based parametrization that ensures unitarity, preserves the Bell scenario structure, and enables efficient optimization (Supplementary Materials~\Cref{app:OptimizationParametrization} for details). 
Concretely, each local measurement setting $x\in\{0,1\}$ at site $i$ is described by a parameter vector $\boldsymbol{\theta}_x^{(i)}$, from which we construct quantum projectors $P_{a|x}^{(i)}(\boldsymbol{\theta}_x^{(i)})$ with outcomes $a\in\{0,1,2\}$. These projectors directly define the parametrized Bell operator $\mathcal{B}\left(\boldsymbol{\theta}\right)$ via Eq.~\eqref{eq:bell_operator}, where $\boldsymbol{\theta}$ collects all $\boldsymbol{\theta}_x^{(i)}$.  

In principle, constructing the Bell operator in~\eqref{eq:bell_operator} would require different measurement parameters $\boldsymbol{\theta}_x^{(i)}$ for each site $i\in\{1,\ldots,n\}$. However, this is too computationally demanding. Instead, exploiting the permutation invariance symmetry inherent to the PIBI \cite{AnnPhys,mekonnen2025invariance}, we assume that the optimal violation can be achieved when all parties share the same measurement pair,
\[
\boldsymbol{\theta}_x^{(i)}=\boldsymbol{\theta}_x \quad \forall\,i\in[n],
\]
reducing the optimization to a global parameter set $\Theta=(\boldsymbol{\theta}_0,\boldsymbol{\theta}_1)$. 
By Schur-Weyl duality, PI Bell operators block-diagonalize in a symmetry-adapted basis, where each block has polynomial size \cite{FadelQuantum2018}. These blocks correspond to irreducible representations of the permutation group. This allows us to simplify the search for optimal measurements by optimizing within each block independently (see Supplementary Materials~\Cref{app:OptimizationSU3irreps} for details). Let's now examine how these symmetries relate to SU(3) Hamiltonians and quantum chaos.

\subsection{SU(3) and quantum chaos} 

The Hilbert space of a three-level system is spanned by the standard basis vectors $\ket{0}=(1,0,0)^T$, $\ket{1}=(0,1,0)^T$ and $\ket{2}=(0,0,1)^T$. Transitions $\ket{\beta}$ to $\ket{\alpha}$ are generated by the ladder operators $\tau_{\alpha\beta}=\ket{\alpha}\!\bra{\beta}$, for $\alpha,\beta\in\{0,1,2\}$.
For $n$ such systems, we define the collective operators $S_{\alpha\beta}=\sum_{i} \tau_{\alpha\beta}^{(i)}$, noting $S_{00}+S_{11}+S_{22}=n \mathbb{I}$. Following standard SU(3) conventions, we define the isospin component $T_3=(S_{00}-S_{11})/2$ and hypercharge $Y=(S_{00}+S_{11}-2S_{22})/2$. 
Irreducible representations of SU(3) (\textit{irreps}) are uniquely characterized by its highest-weight vector $\ket{\mu}$, which is defined as the common eigenstate of $T_3$ an $Y$ with eigenvalues
\begin{equation}\label{eq:T3Ydef}
    T_3\ket{\mu}=\dfrac{p}{2}\ket{\mu} \;,\quad Y\ket{\mu}=\dfrac{p+2q}{3}\ket{\mu} \;,
\end{equation}
and it is annihilated by $S_{01}\ket{\mu}=S_{12}\ket{\mu}=S_{02}\ket{\mu}=0$. Analogously to the spin number $0\leq J\leq n/2$ used to label SU(2) irreps, we use here the pair of non-negative integers $(p, q)$ to label SU(3) irreps. Each irrep has dimension $(1+p)(1+q)(2+p+q)/2$.

SU(3) Hamiltonians, such as the Elliott model describing collective excitations in nuclear physics \cite{Elliott1958}, are valuable tools for probing quantum chaos and the role of symmetry-breaking influencing chaotic dynamics. In their semiclassical limit (large irrep dimension), these models commonly exhibit chaotic behavior \cite{YeffeRMP,GnutzmannJPhysA1999}. 

Quantum chaos is often characterized via statistical properties of energy-level spacings \cite{Haake2018}. Traditionally, a key tool is the nearest-neighbor energy-level spacing distribution (NNSD): NNSDs with Poisson statistics indicate that the energy levels are generally uncorrelated signaling integrability, while NNSDs with Wigner-Dyson statistics signal chaos via ``level repulsion'', as described by random matrix theory.
However, NNSD requires spectrum unfolding, introducing unwanted complexity and parameter dependence. Instead, we primarily use the \textit{ratio of consecutive level spacings} (RCS) \cite{oganesyan2007localization, atas2013distribution}, a robust alternative chaos indicator that avoids unfolding by directly measuring level correlations. For completion, NNSD results are also provided in the Supplementary Material~\cref{app:NNSD}, confirming agreement across both methods.

To evaluate the RCS distribution $P(r)$, we compute $r_l:=\frac{s_l}{s_{l-1}}$ from an ordered spectrum $\{e_0,e_1,\ldots,e_{L-1}\}$ of $L$ energy-levels, where $s_l := e_{l+1}-e_l$ is the nearest-neighbor spacing. Then, the histogram of $\{r_l\}$ yields $P(r)$, which can be fit using the interpolation formula~\cite{karampagia2022ratio}
\begin{eqnarray}
\label{eq:RCSinterpolation}
P(r,\lambda)=C_{\lambda}\frac{(r+r^2)^{\lambda}}{(1+r+\lambda r^2)^{2+\frac{1}{2}\lambda}},
\end{eqnarray}
where $C_{\lambda}=\Gamma\left[\frac{3.72+\lambda}{1.86+\lambda}\right]^{-(10+\lambda)}$ is a normalization constant, $\Gamma$ denotes the gamma function, and $0\leq\lambda\leq 1$ is the RCS parameter which interpolates between Poisson ($\lambda = 0$) and GOE Wigner-Dyson ($\lambda= 1$) distributions (see Fig.~\ref{fig1} for an example).

\begin{figure}[t]
\centering
\includegraphics[width=0.95\linewidth]{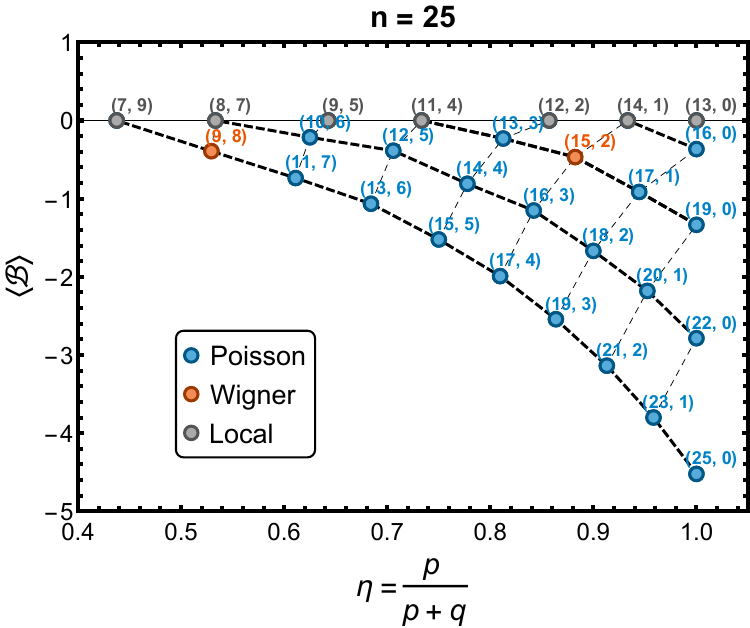}
\caption{
Maximal quantum violation of the PIBI~(\ref{eq:PIBI}) for $n=25$ parties, restricted to the $(p,q)$ irrep subsector of SU(3). The classical bound is $\beta_c=0$, so $\braket{\mathcal{B}}<0$ certifies nonlocality. The parameter $\eta=p/(p+q)$ quantifies the degree of permutation invariance of each irrep, with $\eta=1$ being the fully symmetric case. 
Blue points correspond to irreps $(p,q)$ whose Bell operator exhibits Poisson RCS statistics, signalled by $\lambda = 0$, indicative of integrability. Orange points correspond to irreps for which the fitted RCS parameter is non-zero ($0<\lambda<1$). The histograms of the orange cases are fitted with values $\lambda = 0.11$ for the irrep $(15,2)$ and $\lambda = 0.456$ for $(9,8)$, suggesting a crossover regime between the Poisson and GOE limits, an interpretation further supported by the significant bin weight in their RCS histograms at small spacing values (see Supplementary \cref{fig:RCS_and_NNSD_n25,fig:RCS_and_NNSD_n25_histograms} and Supplementary \cref{tab:rcs_nnsd_n25} for explicit values of all $\lambda$'s obtained and some illustrative histograms). 
Irreps shown in gray do not detect nonlocality and are included only for completeness. Dashed lines connect irreps with same $p+2q$ and $p-q$. The largest violations occur in the fully symmetric sector, as expected from the permutationally invariant structure of the Bell operator \cite{AnnPhys,FadelQuantum2018}.} 
\label{fig2}
\end{figure}

\subsection{PIBI irreducible representations and energy spacing distribution} 

The PIBI~\eqref{eq:PIBI} detects nonlocality in low-energy states of SU(3) Hamiltonians \cite{mullerrigat2024threeoutcome}, relevant for quantum chaos \cite{MeredithPRA1988,GnutzmannJPhysA1999,LipkinNucPhys1965}. Interestingly, the associated Bell operator can be viewed as a Hamiltonian whose low-energy states display nonlocality \cite{TuraPRX2017,FadelQuantum2018}, hinting at a potential link between Bell nonlocality and quantum chaos.

To explore this, we select an SU(3) irrep $(p,q)$ and optimize the measurement parameters $\bm{\theta}$ within this subsector to minimize the Bell operator eigenvalue (\textit{i.e.,} find maximal violation when negative). We then compute the RCS distribution and extract the best-fit $\lambda$ using Eq.~\eqref{eq:RCSinterpolation} (see Fig.~\ref{fig1}). Repeating this across all inequivalent irreps $(p,q)$ allows us compare quantum violations with RCS statistics. 
In Fig.~\ref{fig2}, we summarize the results for $n=25$ parties, plotting quantum violations against $\eta = p/(p+q) \in [0,1]$, a measure of permutation symmetry akin to \cite{GnutzmannJPhysA1999}, with $\eta = 1$ corresponding to fully symmetric irreps. In their work, different $\eta$ lead to distinct classical limits, some chaotic and some not, a feature absent in SU(2). 

A key trend emerges: irreps whose measurement configurations yield maximal Bell violation typically exhibit RCS distributions characteristic of integrable behaviour, namely Poisson statistics ($\lambda = 0$), indicating level clustering. On the other hand, non-optimal or random measurement configurations generically display level repulsion and chaotic spectral characteristics, well described by Wigner-Dyson statistics ($\lambda > 0$). 

The emergence of Poissonian RCS statistics at optimal nonlocal measurements, and Wigner-Dyson statistics otherwise, is consistently observed from $n=8$ (the smallest system size for which our PIBI detects nonlocality) up to $n=32$ (our computational limit). Fig.~\ref{fig2} also includes apparent exceptions (orange points), for which optimal measurement configurations yield non-zero fitted RCS parameters $\lambda > 0$. We interpret these apparent exceptions as finite-size effects arising from the limited Hilbert-space dimension of the corresponding irreps. Furthermore, a closer inspection of these cases (see Supplementary Materials \cref{fig:RCS_and_NNSD_n25_histograms}) suggests that they do not exhibit fully developed level repulsion. In particular, their RCS histograms display a significant non-vanishing weight at near zero spacings $r$, which reveal that these cases lie in a crossover regime between Poisson and Wigner-Dyson statistics rather than genuine chaotic behaviour. This interpretation is consistent with the intermediate RCS fitted values $0<\lambda<1$ observed for these irreps.

By contrast, for random measurement configurations (see Section~\ref{sec:randMeas}), we generically observe substantially larger values of $\lambda$ and RCS distributions closely matching GOE predictions. Taken together, these observations support our interpretation that the apparent deviations are not true counterexamples of the integrable behaviour for optimal measurement settings, but arise from finite-size crossover artifacts due to limited Hilbert-space dimension that vanish in the asymptotic limit.
We therefore conjecture that, in the large-$n$ limit, non-local irreps with optimal measurements that maximally violate the Bell inequality universally approach integrable RCS statistics within this class of Bell operators.

Additionally, we observe remarkable intriguing patterns when plotting maximal quantum violations against $\eta$. 
For instance, irreps with the same hypercharge $Y=p+2q$ (Eq.~\eqref{eq:T3Ydef}) align along well-defined curves (dashed lines in Fig.~\ref{fig2}). These structures suggest deeper analytical relations between the PIBI maximal violations and the degree of permutation symmetry of SU(3) subsectors, which we leave open for future work.

\subsection{Robustness of integrability at maximum quantum violation}
The choice of measurement settings $\bm{\theta}$ is crucial, since it defines the Bell operator and, therefore, its quantum violation and its spectral properties. Building on the observation that restricting to the measurements settings leading to maximal violation generically results in a Bell operator with Poisson RCS distribution, in this section we explore what happens for more general measurement choices. 

In particular, to assess the robustness of the integrable behavior we analyze: i) what happens under random measurement choices; and ii) perturbations around the optimal point. In both cases, we will see in what follows that the integrability trend rapidly disappears. Random measurements generically lead to Wigner-Dyson statistics, signature of quantum chaos, while mild deviations from the optimal measurement settings induce a transition from Poisson to Wigner-Dyson statistics. Moreover, the volume of measurement settings around the optimal yielding Poisson-like RCS distributions shrinks with increasing $n$. Both findings suggest that integrable behavior is indeed a rare, fine-tuned feature of the optimal measurement configurations, rather than a generic property of the Bell operator.

\subsubsection{Random measurement settings}\label{sec:randMeas}
\begin{figure}[th]
\centering
\includegraphics[width=0.9\linewidth]{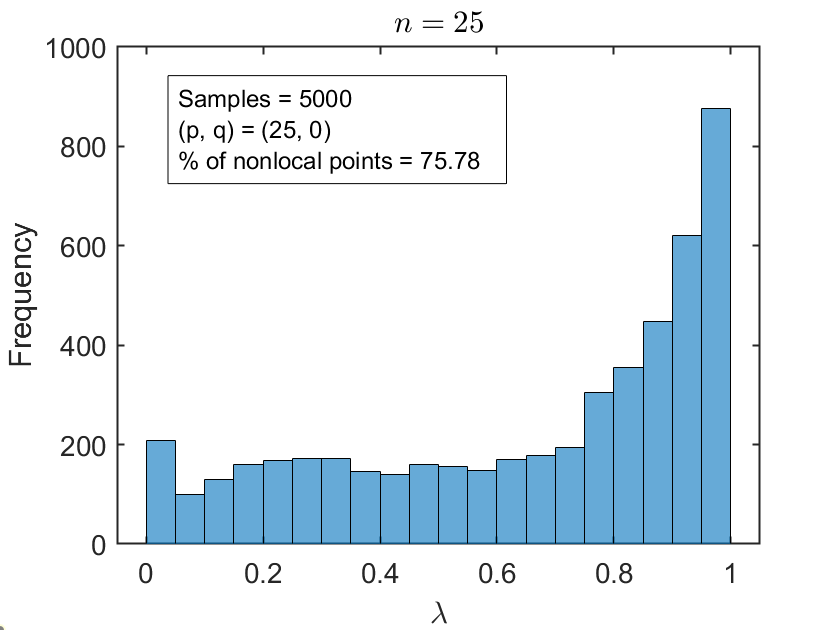}
\caption{Histogram of RCS parameters $\lambda$ resulting from fitting the RCS distribution of the Bell operator constructed from the PIBI~\eqref{eq:PIBI} with random projectors. 
Here $n=25$ and $(p,q)=(25,0)$, \ie the lowest point in Fig.~\ref{fig2}, but other irreps show a similar behaviour despite having a lower fraction of points exhibiting nonlocality.
}
\label{fig3}
\end{figure}

To investigate the generic case, we study here the connection between the violation of PIBI~\eqref{eq:PIBI} and the RCS distribution in the case of pairs of random measurement settings.
For every irrep, we have generated more than $10^3$ random projectors obtained by appropriately sampling matrices from SU(3), and then computed the RCS distribution of the resulting Bell operator. 
We show in Fig.~\ref{fig3} the histogram of the fitted $\lambda$ parameters for the illustrative case $(p,q)=(25,0)$ and $10^4$ samples.

Generically, we observe that the RCS distribution now shows level repulsion ($\lambda > 0$), approaching a Wigner-Dyson distribution with most cases being GOE ($\lambda=\sim 1$), regardless of the quantum violation. Interestingly, sampling over random projectors yields a high percentage of nonlocality detection. We attribute this behavior to the restriction of using identical measurement settings $\bm{\theta}^{(i)}=\bm{\theta}^{(j)}=\bm{\theta}$ for all parties $i,j\in [n]$, which adds a restrictive structure. In an even more general scenario, where each party could set different random measurements, we would expect the departure from Poisson RCS distributions to be even more accentuated. While chaotic behavior in the Bell operator is typically expected when sampling random measurements, this highlights the remarkable property of observing Poisson RCS distribution when using optimal measurements for each irrep. 

To complete the picture, in the Supplementary Material Fig.~\ref{fig:otherRandProj} we show the results for additional representative irreps and also for $n=20$, consistently displaying the same RCS behavior.

\subsubsection{Volume of Poisson-like behavior around the optimal point} 
To quantify the robustness of the connection between Poisson RCS distribution and measurement optimality, we proceed as follows: For a given $n$ and irrep, we start from the Bell operator with optimal measurements and gradually deviate from it by smoothly perturbing the measurement settings until the RCS distribution of the resulting Bell operator stops being Poissonian. This approach allows us to estimate a region of measurement settings that yield a Poisson RCS distribution, whose volume we observe shrinking as $n$ increases.


\begin{figure}[t]
\centering
\includegraphics[width=1\linewidth]{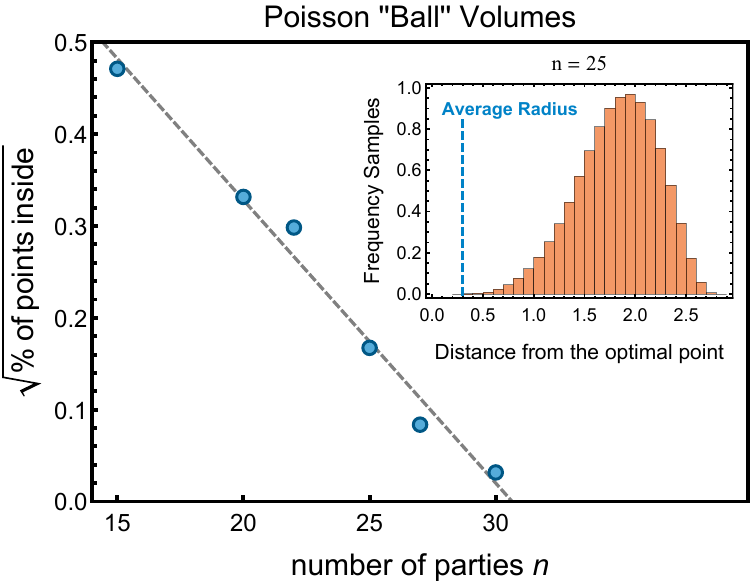}
\caption{Estimated volume of the region of measurement settings  resulting in a Poisson RCS distribution plotted as a function of the number of parties $n$. We observe that the volume decreases as the number of parties $n$ increases. The inset for $n=25$ shows that only a small percentage of the randomly generated observables fall inside the estimated Poissonian region (see the text for details).}
\label{fig4}
\end{figure}

For this analysis, we use the parametrization~\cref{eq:unitaryparametrized} in Methods to generate a random direction $\bm{\theta}^{\text{rand}}\in\mathbb{R}^9$ in the space of measurement settings parameters, with $||\bm{\theta}^{\text{rand}}|| \ll 1$. We then iteratively add $\bm{\theta}^{\text{rand}}$ as a perturbation to the optimal measurements $\bm{\theta}^{\text{opt}}$. At each step, we check the new Bell operator and its RCS distribution.
After a sufficient number of iterations $\alpha\in \mathbb{N}$, we observe the RCS distribution transition from being Poisson ($\lambda \sim 0$) to Wigner-Dyson ($\lambda > 0$). The point at which this transition occurs defines the measurement settings $\tilde{\bm{\theta}}=\bm{\theta}^{\text{opt}}+\alpha\bm{\theta}^{\text{rand}}$.
By repeating this procedure for several random directions $\bm{\theta}^{\text{rand}}_j$ we obtain a set of points in the measurement settings space, centered around $\bm{\theta}^{\textrm{opt}}$, which samples the boundary of the region where the RCS distribution remains Poissonian. 
After having collected enough boundary points $\tilde{\bm{\theta}}$, we use the corresponding projectors to obtain observables $A(\tilde{\bm{\theta}})$ and use the Frobenius distance to compute the average radius $R^{\text{avg}}$ of the region with respect to the optimal settings. 
That is, we compute $R^{\text{avg}}=\sum_{j=1}^s R_j/s$, where $s$ is the number of estimated boundary points we obtained and $R_j=\mathrm{Tr}\left[\left(A(\bm{\theta}^{\text{opt}})- A(\tilde{\bm{\theta}}_j)\right)\left(A(\bm{\theta}^{\text{opt}})- A(\tilde{\bm{\theta}}_j)\right)^\dagger \right]^{1/2}$. 
Therefore, we identify $R^{\text{avg}}$ as the radius of a ball approximating the Poissonian region. Finally, we use a Monte Carlo method to estimate the volume of this region: we generate random observables $A(\bm{\theta})$ and estimate whether they fall within the Poissonian region by comparing the radius $R$ to $R^{\text{avg}}$. Concretely, we check whether their radius $R$ is smaller or larger than the average radius of the region $R^{\text{avg}}$. To ensure that the random observables have been generated uniformly, we construct them from random unitaries $U\in\mathbb{C}^{3\times 3}$ sampled from the Haar measure.


In Fig.~\ref{fig4}, we have estimated the Poissonian regions for each displayed value of $n$ using $10^3$ random directions and $10^5$ random unitaries $U$. We observe that the volumes associated to Poisson RCS distribution diminish as the number of parties increases, suggesting that this volume tends to zero in the asymptotic limit. These results further support our conjecture that the Poisson RCS distribution observed around the point of maximal Bell inequality~\eqref{eq:PIBI} violation is indeed a special case. This observation applies to the maximal violation of each irrep exhibiting nonlocality. 

This analysis reveals the fragile nature of the integrable behavior: it emerges only within a vanishingly small neighborhood around the optimal settings, suggesting that the Poissonian statistics are a singular feature of the optimal configuration.

\subsection{Emergence of Parity Symmetry at Maximal Quantum Violation}

\begin{figure}[t]
\centering
\includegraphics[width=1\linewidth]{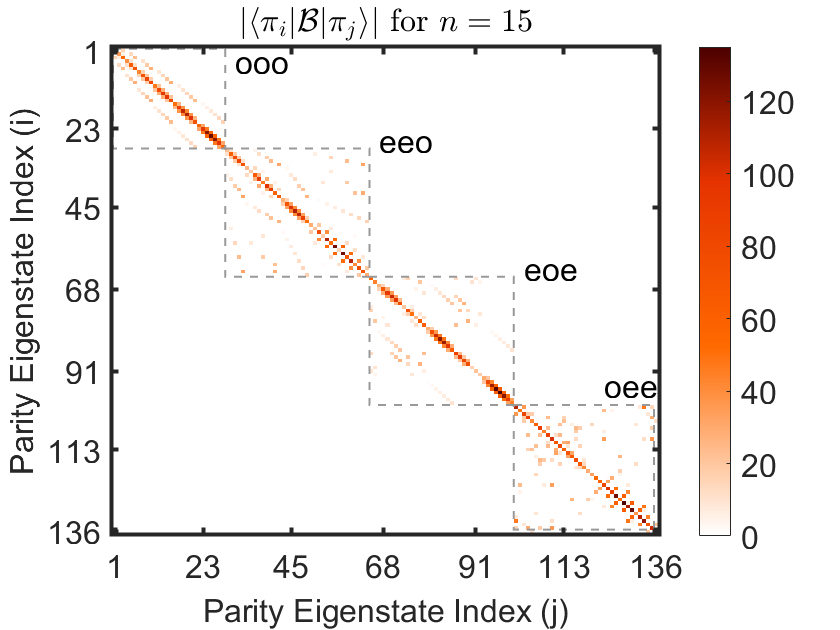}
\caption{Optimal Bell operator $\mathcal{B}$ for $n=15$ qutrits in the parity eigenbasis, revealing a block-diagonal structure with four non-empty parity sectors (for odd $n$ these are $ooo$, $eeo$, $eoe$, and $oee$). Gray boxes enclose the sectors as a visual guide. The color scale indicates matrix element magnitudes; zero entries are shown in white to highlight their sparsity.}
\label{fig:blockdiagonal}
\end{figure}

The Poissonian level statistics observed in Bell operators near maximal quantum violation suggest the presence of an underlying symmetry. Indeed, inspired by~\cite{GrassPRL2013}, we find that for near-optimal measurement settings the Bell operator $\mathcal{B}(\bm{\theta})$ commutes with \emph{parity operators} $\pi_{a\in \{0,1,2\}}$, acting as $\pi_a \ket{n_0, n_1, n_2} = (-1)^{n_a} \ket{n_0, n_1, n_2}$ where $\ket{n_0,n_1,n_2}$ is an $n$-qutrit Dicke state~\cite{ReviewSymmetrics} with $n=n_0+n_1+n_2$. These parity operators partition the symmetric Hilbert space into invariant sectors characterized by whether the number of qutrits $n_a$ at each sublevel $a$ is even ($e$) or odd ($o$). The commutation $[\mathcal{B}(\bm{\theta}^{\textrm{opt}}), \pi_a] = 0$ implies block-diagonalization of $\mathcal{B}(\bm{\theta}^{\textrm{opt}})$ in the Dicke basis (Fig.~\ref{fig:blockdiagonal}). This structure naturally explains the emergence of Poisson statistics: within each block, the Bell operator acts on a reduced effective Hilbert space, and the decoupling between blocks suppresses level repulsion across the full spectrum, resulting in a Poissonian RCS distribution.

Remarkably, the maximally violating eigenstate consistently lies in a specific sector: $eeo$ for odd $n$, and $eee$ for even $n$. This confirms that the Bell operator, when optimized, selects a distinct symmetry sector, indicating the emergence of parity symmetry. 
Therefore, we can conclude that the integrable (Poissonian) spectrum near Bell operators with maximal nonlocality detection also reflects the genuine emergence of a symmetry structure and it is not a numerical artifact. 
This reveals a promising interplay between symmetry, spectral statistics, and Bell nonlocality.

\section{CONCLUSIONS AND DISCUSSION}

We investigated the connection between Bell nonlocality and quantum chaos by introducing a two-input three-outcome Bell inequality for many-body systems, and analyzing the spectral properties of the corresponding Bell operators.
By characterizing the ratio of consecutive level spacings, we found that measurement settings exhibiting maximal Bell nonlocality typically yield Poisson statistics, a signature of integrable dynamics. In contrast, generic (non-optimal) measurements yield Wigner-Dyson statistics, a hallmark of quantum chaos.

These findings suggest that integrability emerges in Bell operators with near-optimal measurement configurations, with even slight perturbations restoring chaotic spectral features. We further uncovered an emergent parity symmetry in near-maximally nonlocal Bell operators, providing a structural explanation for the observed spectral regularity and reinforcing the special nature of these configurations.

Our approach differs fundamentally from previous efforts linking chaos to quantum correlations. Earlier studies typically probe chaos through properties of quantum states such as entanglement growth~\cite{trail2008entanglement}, quantum discord~\cite{madhok2015signatures}, or the violation of Leggett-Garg inequalities in chaotic dynamics~\cite{ramchander2019quantum}, all of which rely on state evolution under a fixed Hamiltonian. By contrast, the framework we present here arises from the structure of the Bell operator itself: its spectral statistics change under different measurement configurations, and integrability emerges from a fine-tuned, symmetry-enhanced point that coincides with maximal Bell violation. This operator-based perspective to probe quantum chaos remains largely unexplored and offers a complementary route to understanding complexity in many-body systems as well as the role quantum nonlocal correlations can play in it.

Our results reveal a link between Bell nonlocality and integrability, within the class of Bell inequalities we consider, providing the first steps toward utilizing Bell inequalities as diagnostic tools for probing quantum chaos. The emergent symmetry and integrable behavior near maximal quantum violation suggest that such configurations may allow simplified characterizations of the underlying quantum state and measurement structure, an essential component for self-testing quantum many-body systems \cite{SupicReview}. Future work could explore whether similar spectral signatures arise in other three-outcome permutationally invariant Bell inequalities~\cite{aloy2024deriving} and extend this analysis to more general multipartite Bell scenarios.

\section{Data availability}
The codes used to generate data for this paper are available at https://github.com/Albert-Aloy/3PIBIs

\section{Acknowledgments and Funding}
These subjects belonged to the beloved areas of the late Fritz Haake, and we dedicate this paper to his memory. 

AA acknowledges support from the Austrian Science Fund (FWF) (projects P 33730-N and 10.55776/PAT2839723) and by the ESQ Discovery programme (Erwin Schr{\"o}dinger Center for Quantum Science \& Technology), hosted by the Austrian Academy of Sciences ({\"O}AW).

JT has received support from the European Union’s Horizon Europe program through the ERC StG FINE-TEA-SQUAD (Grant No. 101040729). Views and opinions expressed are, however, those of the author(s) only and do not necessarily reflect those of the European Union, the European Commission, or the European Space Agency, and neither can they be held responsible for them. J.T. also acknowledges support from the Quantum Delta NL program. This publication is part of the ‘Quantum Inspire – the Dutch Quantum Computer in the Cloud’ project (with project number [NWA.1292.19.194]) of the NWA research program ‘Research on Routes by Consortia (ORC)’, which is funded by the Netherlands Organization for Scientific Research (NWO).

GM and ML acknowledge support from: European Research Council AdG NOQIA; MCIN/AEI (PGC2018-0910.13039/501100011033, CEX2019-000910-S/10.13039/501100011033, Plan National FIDEUA PID2019-106901GB-I00, Plan National STAMEENA PID2022-139099NB, I00, project funded by MCIN/AEI/10.13039/501100011033 and by the “European Union NextGenerationEU/PRTR" (PRTR-C17.I1), FPI); QUANTERA MAQS PCI2019-111828-2); QUANTERA DYNAMITE PCI2022-132919, QuantERA II Programme co-funded by European Union’s Horizon 2020 program under Grant Agreement No 101017733); Ministry for Digital Transformation and of Civil Service of the Spanish Government through the QUANTUM ENIA project call - Quantum Spain project, and by the European Union through the Recovery, Transformation and Resilience Plan - NextGenerationEU within the framework of the Digital Spain 2026 Agenda; Fundació Cellex; Fundació Mir-Puig; Generalitat de Catalunya (European Social Fund FEDER and CERCA program, AGAUR Grant No. 2021 SGR 01452, QuantumCAT \ U16-011424, co-funded by ERDF Operational Program of Catalonia 2014-2020); Barcelona Supercomputing Center MareNostrum (FI-2023-3-0024); Funded by the European Union. Views and opinions expressed are however those of the author(s) only and do not necessarily reflect those of the European Union, European Commission, European Climate, Infrastructure and Environment Executive Agency (CINEA), or any other granting authority. Neither the European Union nor any granting authority can be held responsible for them (HORIZON-CL4-2022-QUANTUM-02-SGA PASQuanS2.1, 101113690, EU Horizon 2020 FET-OPEN OPTOlogic, Grant No 899794), EU Horizon Europe Program (This project has received funding from the European Union’s Horizon Europe research and innovation program under grant agreement No 101080086 NeQSTGrant Agreement 101080086 — NeQST); ICFO Internal “QuantumGaudi” project; European Union’s Horizon 2020 program under the Marie Sklodowska-Curie grant agreement No 847648; “La Caixa” Junior Leaders fellowships, La Caixa” Foundation (ID 100010434): CF/BQ/PR23/11980043.

MF was supported by the Branco Weiss Fellowship -- Society in Science, administered by the ETH Z\"{u}rich.

\bibliographystyle{apsrev4-2} 
\bibliography{ChaosBib.bib}

\begin{thebibliography}{49}%
\makeatletter
\providecommand \@ifxundefined [1]{%
 \@ifx{#1\undefined}
}%
\providecommand \@ifnum [1]{%
 \ifnum #1\expandafter \@firstoftwo
 \else \expandafter \@secondoftwo
 \fi
}%
\providecommand \@ifx [1]{%
 \ifx #1\expandafter \@firstoftwo
 \else \expandafter \@secondoftwo
 \fi
}%
\providecommand \natexlab [1]{#1}%
\providecommand \enquote  [1]{``#1''}%
\providecommand \bibnamefont  [1]{#1}%
\providecommand \bibfnamefont [1]{#1}%
\providecommand \citenamefont [1]{#1}%
\providecommand \href@noop [0]{\@secondoftwo}%
\providecommand \href [0]{\begingroup \@sanitize@url \@href}%
\providecommand \@href[1]{\@@startlink{#1}\@@href}%
\providecommand \@@href[1]{\endgroup#1\@@endlink}%
\providecommand \@sanitize@url [0]{\catcode `\\12\catcode `\$12\catcode
  `\&12\catcode `\#12\catcode `\^12\catcode `\_12\catcode `\%12\relax}%
\providecommand \@@startlink[1]{}%
\providecommand \@@endlink[0]{}%
\providecommand \url  [0]{\begingroup\@sanitize@url \@url }%
\providecommand \@url [1]{\endgroup\@href {#1}{\urlprefix }}%
\providecommand \urlprefix  [0]{URL }%
\providecommand \Eprint [0]{\href }%
\providecommand \doibase [0]{https://doi.org/}%
\providecommand \selectlanguage [0]{\@gobble}%
\providecommand \bibinfo  [0]{\@secondoftwo}%
\providecommand \bibfield  [0]{\@secondoftwo}%
\providecommand \translation [1]{[#1]}%
\providecommand \BibitemOpen [0]{}%
\providecommand \bibitemStop [0]{}%
\providecommand \bibitemNoStop [0]{.\EOS\space}%
\providecommand \EOS [0]{\spacefactor3000\relax}%
\providecommand \BibitemShut  [1]{\csname bibitem#1\endcsname}%
\let\auto@bib@innerbib\@empty
\bibitem [{\citenamefont {Bell}(1964)}]{Bell1964}%
  \BibitemOpen
  \bibfield  {author} {\bibinfo {author} {\bibfnamefont {J.~S.}\ \bibnamefont
  {Bell}},\ }\href@noop {} {\bibfield  {journal} {\bibinfo  {journal}
  {Physics}\ }\textbf {\bibinfo {volume} {1}},\ \bibinfo {pages} {195}
  (\bibinfo {year} {1964})}\BibitemShut {NoStop}%
\bibitem [{\citenamefont {Brunner}\ \emph {et~al.}(2014)\citenamefont
  {Brunner}, \citenamefont {Cavalcanti}, \citenamefont {Pironio}, \citenamefont
  {Scarani},\ and\ \citenamefont {Wehner}}]{BrunnerRMP2014}%
  \BibitemOpen
  \bibfield  {author} {\bibinfo {author} {\bibfnamefont {N.}~\bibnamefont
  {Brunner}}, \bibinfo {author} {\bibfnamefont {D.}~\bibnamefont {Cavalcanti}},
  \bibinfo {author} {\bibfnamefont {S.}~\bibnamefont {Pironio}}, \bibinfo
  {author} {\bibfnamefont {V.}~\bibnamefont {Scarani}},\ and\ \bibinfo {author}
  {\bibfnamefont {S.}~\bibnamefont {Wehner}},\ }\href
  {https://doi.org/10.1103/RevModPhys.86.419} {\bibfield  {journal} {\bibinfo
  {journal} {Rev. Mod. Phys.}\ }\textbf {\bibinfo {volume} {86}},\ \bibinfo
  {pages} {419} (\bibinfo {year} {2014})}\BibitemShut {NoStop}%
\bibitem [{\citenamefont {Fadel}\ and\ \citenamefont
  {Tura}(2018)}]{FadelQuantum2018}%
  \BibitemOpen
  \bibfield  {author} {\bibinfo {author} {\bibfnamefont {M.}~\bibnamefont
  {Fadel}}\ and\ \bibinfo {author} {\bibfnamefont {J.}~\bibnamefont {Tura}},\
  }\href {https://doi.org/10.22331/q-2018-11-19-107} {\bibfield  {journal}
  {\bibinfo  {journal} {{Quantum}}\ }\textbf {\bibinfo {volume} {2}},\ \bibinfo
  {pages} {107} (\bibinfo {year} {2018})}\BibitemShut {NoStop}%
\bibitem [{\citenamefont {Piga}\ \emph {et~al.}(2019)\citenamefont {Piga},
  \citenamefont {Aloy}, \citenamefont {Lewenstein},\ and\ \citenamefont
  {Fr{\'{e}}rot}}]{PigaPRL2019}%
  \BibitemOpen
  \bibfield  {author} {\bibinfo {author} {\bibfnamefont {A.}~\bibnamefont
  {Piga}}, \bibinfo {author} {\bibfnamefont {A.}~\bibnamefont {Aloy}}, \bibinfo
  {author} {\bibfnamefont {M.}~\bibnamefont {Lewenstein}},\ and\ \bibinfo
  {author} {\bibfnamefont {I.}~\bibnamefont {Fr{\'{e}}rot}},\ }\href
  {10.1103/physrevlett.123.170604} {\bibfield  {journal} {\bibinfo  {journal}
  {Physical Review Letters}\ }\textbf {\bibinfo {volume} {123}} (\bibinfo
  {year} {2019})}\BibitemShut {NoStop}%
\bibitem [{\citenamefont {Fr\"owis}\ \emph {et~al.}(2019)\citenamefont
  {Fr\"owis}, \citenamefont {Fadel}, \citenamefont {Treutlein}, \citenamefont
  {Gisin},\ and\ \citenamefont {Brunner}}]{FroewisArxiv2018}%
  \BibitemOpen
  \bibfield  {author} {\bibinfo {author} {\bibfnamefont {F.}~\bibnamefont
  {Fr\"owis}}, \bibinfo {author} {\bibfnamefont {M.}~\bibnamefont {Fadel}},
  \bibinfo {author} {\bibfnamefont {P.}~\bibnamefont {Treutlein}}, \bibinfo
  {author} {\bibfnamefont {N.}~\bibnamefont {Gisin}},\ and\ \bibinfo {author}
  {\bibfnamefont {N.}~\bibnamefont {Brunner}},\ }\href
  {https://doi.org/10.1103/PhysRevA.99.040101} {\bibfield  {journal} {\bibinfo
  {journal} {Phys. Rev. A}\ }\textbf {\bibinfo {volume} {99}},\ \bibinfo
  {pages} {040101} (\bibinfo {year} {2019})}\BibitemShut {NoStop}%
\bibitem [{\citenamefont {Chazelle}(1993)}]{ChazelleDCG1993}%
  \BibitemOpen
  \bibfield  {author} {\bibinfo {author} {\bibfnamefont {B.}~\bibnamefont
  {Chazelle}},\ }\href {https://doi.org/10.1007/bf02573985} {\bibfield
  {journal} {\bibinfo  {journal} {Discrete {\&} Computational Geometry}\
  }\textbf {\bibinfo {volume} {10}},\ \bibinfo {pages} {377} (\bibinfo {year}
  {1993})}\BibitemShut {NoStop}%
\bibitem [{\citenamefont {Tura}\ \emph {et~al.}(2014)\citenamefont {Tura},
  \citenamefont {Augusiak}, \citenamefont {Sainz}, \citenamefont {V{\'e}rtesi},
  \citenamefont {Lewenstein},\ and\ \citenamefont {Ac{\'\i}n}}]{SciencePaper}%
  \BibitemOpen
  \bibfield  {author} {\bibinfo {author} {\bibfnamefont {J.}~\bibnamefont
  {Tura}}, \bibinfo {author} {\bibfnamefont {R.}~\bibnamefont {Augusiak}},
  \bibinfo {author} {\bibfnamefont {A.~B.}\ \bibnamefont {Sainz}}, \bibinfo
  {author} {\bibfnamefont {T.}~\bibnamefont {V{\'e}rtesi}}, \bibinfo {author}
  {\bibfnamefont {M.}~\bibnamefont {Lewenstein}},\ and\ \bibinfo {author}
  {\bibfnamefont {A.}~\bibnamefont {Ac{\'\i}n}},\ }\href
  {https://doi.org/10.1126/science.1247715} {\bibfield  {journal} {\bibinfo
  {journal} {Science}\ }\textbf {\bibinfo {volume} {344}},\ \bibinfo {pages}
  {1256} (\bibinfo {year} {2014})}\BibitemShut {NoStop}%
\bibitem [{\citenamefont {Wagner}\ \emph {et~al.}(2017)\citenamefont {Wagner},
  \citenamefont {Schmied}, \citenamefont {Fadel}, \citenamefont {Treutlein},
  \citenamefont {Sangouard},\ and\ \citenamefont {Bancal}}]{WagnerPRL2017}%
  \BibitemOpen
  \bibfield  {author} {\bibinfo {author} {\bibfnamefont {S.}~\bibnamefont
  {Wagner}}, \bibinfo {author} {\bibfnamefont {R.}~\bibnamefont {Schmied}},
  \bibinfo {author} {\bibfnamefont {M.}~\bibnamefont {Fadel}}, \bibinfo
  {author} {\bibfnamefont {P.}~\bibnamefont {Treutlein}}, \bibinfo {author}
  {\bibfnamefont {N.}~\bibnamefont {Sangouard}},\ and\ \bibinfo {author}
  {\bibfnamefont {J.-D.}\ \bibnamefont {Bancal}},\ }\href
  {https://doi.org/10.1103/physrevlett.119.170403} {\bibfield  {journal}
  {\bibinfo  {journal} {Physical Review Letters}\ }\textbf {\bibinfo {volume}
  {119}},\ \bibinfo {pages} {170403} (\bibinfo {year} {2017})}\BibitemShut
  {NoStop}%
\bibitem [{\citenamefont {Baccari}\ \emph {et~al.}(2019)\citenamefont
  {Baccari}, \citenamefont {Tura}, \citenamefont {Fadel}, \citenamefont {Aloy},
  \citenamefont {Bancal}, \citenamefont {Sangouard}, \citenamefont
  {Lewenstein}, \citenamefont {Ac{\'{\i}}n},\ and\ \citenamefont
  {Augusiak}}]{BaccariPRA2019}%
  \BibitemOpen
  \bibfield  {author} {\bibinfo {author} {\bibfnamefont {F.}~\bibnamefont
  {Baccari}}, \bibinfo {author} {\bibfnamefont {J.}~\bibnamefont {Tura}},
  \bibinfo {author} {\bibfnamefont {M.}~\bibnamefont {Fadel}}, \bibinfo
  {author} {\bibfnamefont {A.}~\bibnamefont {Aloy}}, \bibinfo {author}
  {\bibfnamefont {J.-D.}\ \bibnamefont {Bancal}}, \bibinfo {author}
  {\bibfnamefont {N.}~\bibnamefont {Sangouard}}, \bibinfo {author}
  {\bibfnamefont {M.}~\bibnamefont {Lewenstein}}, \bibinfo {author}
  {\bibfnamefont {A.}~\bibnamefont {Ac{\'{\i}}n}},\ and\ \bibinfo {author}
  {\bibfnamefont {R.}~\bibnamefont {Augusiak}},\ }\href
  {10.1103/physreva.100.022121} {\bibfield  {journal} {\bibinfo  {journal}
  {Physical Review A}\ }\textbf {\bibinfo {volume} {100}} (\bibinfo {year}
  {2019})}\BibitemShut {NoStop}%
\bibitem [{\citenamefont {Fadel}\ and\ \citenamefont
  {Tura}(2017)}]{FadelPRL2017}%
  \BibitemOpen
  \bibfield  {author} {\bibinfo {author} {\bibfnamefont {M.}~\bibnamefont
  {Fadel}}\ and\ \bibinfo {author} {\bibfnamefont {J.}~\bibnamefont {Tura}},\
  }\href {https://doi.org/10.1103/PhysRevLett.119.230402} {\bibfield  {journal}
  {\bibinfo  {journal} {Phys. Rev. Lett.}\ }\textbf {\bibinfo {volume} {119}},\
  \bibinfo {pages} {230402} (\bibinfo {year} {2017})}\BibitemShut {NoStop}%
\bibitem [{\citenamefont {Guo}\ \emph {et~al.}(2023)\citenamefont {Guo},
  \citenamefont {Tura}, \citenamefont {He},\ and\ \citenamefont
  {Fadel}}]{GuoPRL23}%
  \BibitemOpen
  \bibfield  {author} {\bibinfo {author} {\bibfnamefont {J.}~\bibnamefont
  {Guo}}, \bibinfo {author} {\bibfnamefont {J.}~\bibnamefont {Tura}}, \bibinfo
  {author} {\bibfnamefont {Q.}~\bibnamefont {He}},\ and\ \bibinfo {author}
  {\bibfnamefont {M.}~\bibnamefont {Fadel}},\ }\href
  {https://doi.org/10.1103/PhysRevLett.131.070201} {\bibfield  {journal}
  {\bibinfo  {journal} {Phys. Rev. Lett.}\ }\textbf {\bibinfo {volume} {131}},\
  \bibinfo {pages} {070201} (\bibinfo {year} {2023})}\BibitemShut {NoStop}%
\bibitem [{\citenamefont {Schmied}\ \emph {et~al.}(2016)\citenamefont
  {Schmied}, \citenamefont {Bancal}, \citenamefont {Allard}, \citenamefont
  {Fadel}, \citenamefont {Scarani}, \citenamefont {Treutlein},\ and\
  \citenamefont {Sangouard}}]{SchmiedScience2016}%
  \BibitemOpen
  \bibfield  {author} {\bibinfo {author} {\bibfnamefont {R.}~\bibnamefont
  {Schmied}}, \bibinfo {author} {\bibfnamefont {J.-D.}\ \bibnamefont {Bancal}},
  \bibinfo {author} {\bibfnamefont {B.}~\bibnamefont {Allard}}, \bibinfo
  {author} {\bibfnamefont {M.}~\bibnamefont {Fadel}}, \bibinfo {author}
  {\bibfnamefont {V.}~\bibnamefont {Scarani}}, \bibinfo {author} {\bibfnamefont
  {P.}~\bibnamefont {Treutlein}},\ and\ \bibinfo {author} {\bibfnamefont
  {N.}~\bibnamefont {Sangouard}},\ }\href
  {https://doi.org/10.1126/science.aad8665} {\bibfield  {journal} {\bibinfo
  {journal} {Science}\ }\textbf {\bibinfo {volume} {352}},\ \bibinfo {pages}
  {441} (\bibinfo {year} {2016})}\BibitemShut {NoStop}%
\bibitem [{\citenamefont {Engelsen}\ \emph {et~al.}(2017)\citenamefont
  {Engelsen}, \citenamefont {Krishnakumar}, \citenamefont {Hosten},\ and\
  \citenamefont {Kasevich}}]{EngelsenPRL2017}%
  \BibitemOpen
  \bibfield  {author} {\bibinfo {author} {\bibfnamefont {N.~J.}\ \bibnamefont
  {Engelsen}}, \bibinfo {author} {\bibfnamefont {R.}~\bibnamefont
  {Krishnakumar}}, \bibinfo {author} {\bibfnamefont {O.}~\bibnamefont
  {Hosten}},\ and\ \bibinfo {author} {\bibfnamefont {M.~A.}\ \bibnamefont
  {Kasevich}},\ }\href {https://doi.org/10.1103/PhysRevLett.118.140401}
  {\bibfield  {journal} {\bibinfo  {journal} {Phys. Rev. Lett.}\ }\textbf
  {\bibinfo {volume} {118}},\ \bibinfo {pages} {140401} (\bibinfo {year}
  {2017})}\BibitemShut {NoStop}%
\bibitem [{\citenamefont {Law}\ \emph {et~al.}(1998)\citenamefont {Law},
  \citenamefont {Pu},\ and\ \citenamefont {Bigelow}}]{LawPRL98}%
  \BibitemOpen
  \bibfield  {author} {\bibinfo {author} {\bibfnamefont {C.~K.}\ \bibnamefont
  {Law}}, \bibinfo {author} {\bibfnamefont {H.}~\bibnamefont {Pu}},\ and\
  \bibinfo {author} {\bibfnamefont {N.~P.}\ \bibnamefont {Bigelow}},\ }\href
  {https://doi.org/10.1103/PhysRevLett.81.5257} {\bibfield  {journal} {\bibinfo
   {journal} {Phys. Rev. Lett.}\ }\textbf {\bibinfo {volume} {81}},\ \bibinfo
  {pages} {5257} (\bibinfo {year} {1998})}\BibitemShut {NoStop}%
\bibitem [{\citenamefont {Hamley}\ \emph {et~al.}(2012)\citenamefont {Hamley},
  \citenamefont {Gerving}, \citenamefont {Hoang}, \citenamefont {Bookjans},\
  and\ \citenamefont {Chapman}}]{HamleyNat12}%
  \BibitemOpen
  \bibfield  {author} {\bibinfo {author} {\bibfnamefont {C.~D.}\ \bibnamefont
  {Hamley}}, \bibinfo {author} {\bibfnamefont {C.~S.}\ \bibnamefont {Gerving}},
  \bibinfo {author} {\bibfnamefont {T.~M.}\ \bibnamefont {Hoang}}, \bibinfo
  {author} {\bibfnamefont {E.~M.}\ \bibnamefont {Bookjans}},\ and\ \bibinfo
  {author} {\bibfnamefont {M.~S.}\ \bibnamefont {Chapman}},\ }\href
  {https://doi.org/10.1038/nphys2245} {\bibfield  {journal} {\bibinfo
  {journal} {Nature Physics}\ }\textbf {\bibinfo {volume} {8}},\ \bibinfo
  {pages} {305} (\bibinfo {year} {2012})}\BibitemShut {NoStop}%
\bibitem [{\citenamefont {Kitzinger}\ \emph {et~al.}(2021)\citenamefont
  {Kitzinger}, \citenamefont {Meng}, \citenamefont {Fadel}, \citenamefont
  {Ivannikov}, \citenamefont {Nemoto}, \citenamefont {Munro},\ and\
  \citenamefont {Byrnes}}]{KitzingerPRA21}%
  \BibitemOpen
  \bibfield  {author} {\bibinfo {author} {\bibfnamefont {J.}~\bibnamefont
  {Kitzinger}}, \bibinfo {author} {\bibfnamefont {X.}~\bibnamefont {Meng}},
  \bibinfo {author} {\bibfnamefont {M.}~\bibnamefont {Fadel}}, \bibinfo
  {author} {\bibfnamefont {V.}~\bibnamefont {Ivannikov}}, \bibinfo {author}
  {\bibfnamefont {K.}~\bibnamefont {Nemoto}}, \bibinfo {author} {\bibfnamefont
  {W.~J.}\ \bibnamefont {Munro}},\ and\ \bibinfo {author} {\bibfnamefont
  {T.}~\bibnamefont {Byrnes}},\ }\href
  {https://doi.org/10.1103/PhysRevA.104.043323} {\bibfield  {journal} {\bibinfo
   {journal} {Phys. Rev. A}\ }\textbf {\bibinfo {volume} {104}},\ \bibinfo
  {pages} {043323} (\bibinfo {year} {2021})}\BibitemShut {NoStop}%
\bibitem [{\citenamefont {Luo}\ \emph {et~al.}(2017)\citenamefont {Luo},
  \citenamefont {Zou}, \citenamefont {Wu}, \citenamefont {Liu}, \citenamefont
  {Han}, \citenamefont {Tey},\ and\ \citenamefont {You}}]{OurFriends}%
  \BibitemOpen
  \bibfield  {author} {\bibinfo {author} {\bibfnamefont {X.-Y.}\ \bibnamefont
  {Luo}}, \bibinfo {author} {\bibfnamefont {Y.-Q.}\ \bibnamefont {Zou}},
  \bibinfo {author} {\bibfnamefont {L.-N.}\ \bibnamefont {Wu}}, \bibinfo
  {author} {\bibfnamefont {Q.}~\bibnamefont {Liu}}, \bibinfo {author}
  {\bibfnamefont {M.-F.}\ \bibnamefont {Han}}, \bibinfo {author} {\bibfnamefont
  {M.~K.}\ \bibnamefont {Tey}},\ and\ \bibinfo {author} {\bibfnamefont
  {L.}~\bibnamefont {You}},\ }\href {https://doi.org/10.1126/science.aag1106}
  {\bibfield  {journal} {\bibinfo  {journal} {Science}\ }\textbf {\bibinfo
  {volume} {355}},\ \bibinfo {pages} {620} (\bibinfo {year}
  {2017})}\BibitemShut {NoStop}%
\bibitem [{\citenamefont {Haldane}(1983{\natexlab{a}})}]{haldane1983continuum}%
  \BibitemOpen
  \bibfield  {author} {\bibinfo {author} {\bibfnamefont {F.~D.~M.}\
  \bibnamefont {Haldane}},\ }\href
  {https://doi.org/10.1016/0375-9601(83)90631-X} {\bibfield  {journal}
  {\bibinfo  {journal} {Physics Letters A}\ }\textbf {\bibinfo {volume} {93}},\
  \bibinfo {pages} {464} (\bibinfo {year} {1983}{\natexlab{a}})}\BibitemShut
  {NoStop}%
\bibitem [{\citenamefont {Haldane}(1983{\natexlab{b}})}]{HaldanePRL83}%
  \BibitemOpen
  \bibfield  {author} {\bibinfo {author} {\bibfnamefont {F.~D.~M.}\
  \bibnamefont {Haldane}},\ }\href
  {https://doi.org/10.1103/PhysRevLett.50.1153} {\bibfield  {journal} {\bibinfo
   {journal} {Phys. Rev. Lett.}\ }\textbf {\bibinfo {volume} {50}},\ \bibinfo
  {pages} {1153} (\bibinfo {year} {1983}{\natexlab{b}})}\BibitemShut {NoStop}%
\bibitem [{\citenamefont {Affleck}\ \emph {et~al.}(1987)\citenamefont
  {Affleck}, \citenamefont {Kennedy}, \citenamefont {Lieb},\ and\ \citenamefont
  {Tasaki}}]{AKLTPRL87}%
  \BibitemOpen
  \bibfield  {author} {\bibinfo {author} {\bibfnamefont {I.}~\bibnamefont
  {Affleck}}, \bibinfo {author} {\bibfnamefont {T.}~\bibnamefont {Kennedy}},
  \bibinfo {author} {\bibfnamefont {E.~H.}\ \bibnamefont {Lieb}},\ and\
  \bibinfo {author} {\bibfnamefont {H.}~\bibnamefont {Tasaki}},\ }\href
  {https://doi.org/10.1103/PhysRevLett.59.799} {\bibfield  {journal} {\bibinfo
  {journal} {Phys. Rev. Lett.}\ }\textbf {\bibinfo {volume} {59}},\ \bibinfo
  {pages} {799} (\bibinfo {year} {1987})}\BibitemShut {NoStop}%
\bibitem [{\citenamefont {Elliott}(1958)}]{Elliott1958}%
  \BibitemOpen
  \bibfield  {author} {\bibinfo {author} {\bibfnamefont {J.~P.}\ \bibnamefont
  {Elliott}},\ }\href@noop {} {\bibfield  {journal} {\bibinfo  {journal}
  {Proceedings of the Royal Society of London. Series A. Mathematical and
  Physical Sciences}\ }\textbf {\bibinfo {volume} {245}},\ \bibinfo {pages}
  {128} (\bibinfo {year} {1958})}\BibitemShut {NoStop}%
\bibitem [{\citenamefont {Fishman}\ \emph {et~al.}(1982)\citenamefont
  {Fishman}, \citenamefont {Grempel},\ and\ \citenamefont
  {Prange}}]{KickedRotator}%
  \BibitemOpen
  \bibfield  {author} {\bibinfo {author} {\bibfnamefont {S.}~\bibnamefont
  {Fishman}}, \bibinfo {author} {\bibfnamefont {D.~R.}\ \bibnamefont
  {Grempel}},\ and\ \bibinfo {author} {\bibfnamefont {R.~E.}\ \bibnamefont
  {Prange}},\ }\href {https://doi.org/10.1103/PhysRevLett.49.509} {\bibfield
  {journal} {\bibinfo  {journal} {Phys. Rev. Lett.}\ }\textbf {\bibinfo
  {volume} {49}},\ \bibinfo {pages} {509} (\bibinfo {year} {1982})}\BibitemShut
  {NoStop}%
\bibitem [{\citenamefont {Haake}\ \emph {et~al.}(1987)\citenamefont {Haake},
  \citenamefont {Ku{\'s}},\ and\ \citenamefont {Scharf}}]{HaakeKickedTop}%
  \BibitemOpen
  \bibfield  {author} {\bibinfo {author} {\bibfnamefont {F.}~\bibnamefont
  {Haake}}, \bibinfo {author} {\bibfnamefont {M.}~\bibnamefont {Ku{\'s}}},\
  and\ \bibinfo {author} {\bibfnamefont {R.}~\bibnamefont {Scharf}},\ }\href
  {https://doi.org/10.1007/BF01303727} {\bibfield  {journal} {\bibinfo
  {journal} {Zeitschrift f{\"u}r Physik B Condensed Matter}\ }\textbf {\bibinfo
  {volume} {65}},\ \bibinfo {pages} {381} (\bibinfo {year} {1987})}\BibitemShut
  {NoStop}%
\bibitem [{\citenamefont {Bohigas}\ \emph {et~al.}(1984)\citenamefont
  {Bohigas}, \citenamefont {Giannoni},\ and\ \citenamefont
  {Schmit}}]{bohigas1984characterization}%
  \BibitemOpen
  \bibfield  {author} {\bibinfo {author} {\bibfnamefont {O.}~\bibnamefont
  {Bohigas}}, \bibinfo {author} {\bibfnamefont {M.~J.}\ \bibnamefont
  {Giannoni}},\ and\ \bibinfo {author} {\bibfnamefont {C.}~\bibnamefont
  {Schmit}},\ }\href {https://doi.org/10.1103/PhysRevLett.52.1} {\bibfield
  {journal} {\bibinfo  {journal} {Phys. Rev. Lett.}\ }\textbf {\bibinfo
  {volume} {52}},\ \bibinfo {pages} {1} (\bibinfo {year} {1984})}\BibitemShut
  {NoStop}%
\bibitem [{\citenamefont {Mehta}(2004)}]{mehta2004random}%
  \BibitemOpen
  \bibfield  {author} {\bibinfo {author} {\bibfnamefont {M.~L.}\ \bibnamefont
  {Mehta}},\ }\href@noop {} {\emph {\bibinfo {title} {Random matrices}}}\
  (\bibinfo  {publisher} {Elsevier},\ \bibinfo {year} {2004})\BibitemShut
  {NoStop}%
\bibitem [{\citenamefont {Haake}\ \emph {et~al.}(2018)\citenamefont {Haake},
  \citenamefont {Gnutzmann},\ and\ \citenamefont {Kuś}}]{Haake2018}%
  \BibitemOpen
  \bibfield  {author} {\bibinfo {author} {\bibfnamefont {F.}~\bibnamefont
  {Haake}}, \bibinfo {author} {\bibfnamefont {S.}~\bibnamefont {Gnutzmann}},\
  and\ \bibinfo {author} {\bibfnamefont {M.}~\bibnamefont {Kuś}},\ }\href
  {https://doi.org/10.1007/978-3-319-97580-1} {\emph {\bibinfo {title} {Quantum
  Signatures of Chaos}}}\ (\bibinfo  {publisher} {Springer International
  Publishing},\ \bibinfo {year} {2018})\BibitemShut {NoStop}%
\bibitem [{\citenamefont {Aloy}\ \emph {et~al.}(2024)\citenamefont {Aloy},
  \citenamefont {M{\"u}ller-Rigat}, \citenamefont {Tura},\ and\ \citenamefont
  {Fadel}}]{aloy2024deriving}%
  \BibitemOpen
  \bibfield  {author} {\bibinfo {author} {\bibfnamefont {A.}~\bibnamefont
  {Aloy}}, \bibinfo {author} {\bibfnamefont {G.}~\bibnamefont
  {M{\"u}ller-Rigat}}, \bibinfo {author} {\bibfnamefont {J.}~\bibnamefont
  {Tura}},\ and\ \bibinfo {author} {\bibfnamefont {M.}~\bibnamefont {Fadel}},\
  }\href@noop {} {\bibfield  {journal} {\bibinfo  {journal} {Entropy}\ }\textbf
  {\bibinfo {volume} {26}},\ \bibinfo {pages} {816} (\bibinfo {year}
  {2024})}\BibitemShut {NoStop}%
\bibitem [{\citenamefont {Einstein}\ \emph {et~al.}(1935)\citenamefont
  {Einstein}, \citenamefont {Podolsky},\ and\ \citenamefont {Rosen}}]{EPR35}%
  \BibitemOpen
  \bibfield  {author} {\bibinfo {author} {\bibfnamefont {A.}~\bibnamefont
  {Einstein}}, \bibinfo {author} {\bibfnamefont {B.}~\bibnamefont {Podolsky}},\
  and\ \bibinfo {author} {\bibfnamefont {N.}~\bibnamefont {Rosen}},\ }\href
  {https://doi.org/10.1103/PhysRev.47.777} {\bibfield  {journal} {\bibinfo
  {journal} {Phys. Rev.}\ }\textbf {\bibinfo {volume} {47}},\ \bibinfo {pages}
  {777} (\bibinfo {year} {1935})}\BibitemShut {NoStop}%
\bibitem [{\citenamefont {Tura}\ \emph {et~al.}(2017)\citenamefont {Tura},
  \citenamefont {De~las Cuevas}, \citenamefont {Augusiak}, \citenamefont
  {Lewenstein}, \citenamefont {Ac\'{\i}n},\ and\ \citenamefont
  {Cirac}}]{TuraPRX2017}%
  \BibitemOpen
  \bibfield  {author} {\bibinfo {author} {\bibfnamefont {J.}~\bibnamefont
  {Tura}}, \bibinfo {author} {\bibfnamefont {G.}~\bibnamefont {De~las Cuevas}},
  \bibinfo {author} {\bibfnamefont {R.}~\bibnamefont {Augusiak}}, \bibinfo
  {author} {\bibfnamefont {M.}~\bibnamefont {Lewenstein}}, \bibinfo {author}
  {\bibfnamefont {A.}~\bibnamefont {Ac\'{\i}n}},\ and\ \bibinfo {author}
  {\bibfnamefont {J.~I.}\ \bibnamefont {Cirac}},\ }\href
  {https://doi.org/10.1103/PhysRevX.7.021005} {\bibfield  {journal} {\bibinfo
  {journal} {Phys. Rev. X}\ }\textbf {\bibinfo {volume} {7}},\ \bibinfo {pages}
  {021005} (\bibinfo {year} {2017})}\BibitemShut {NoStop}%
\bibitem [{\citenamefont {Tura}\ \emph {et~al.}(2015)\citenamefont {Tura},
  \citenamefont {Augusiak}, \citenamefont {Sainz}, \citenamefont {L{\"u}cke},
  \citenamefont {Klempt}, \citenamefont {Lewenstein},\ and\ \citenamefont
  {Ac{\'i}n}}]{AnnPhys}%
  \BibitemOpen
  \bibfield  {author} {\bibinfo {author} {\bibfnamefont {J.}~\bibnamefont
  {Tura}}, \bibinfo {author} {\bibfnamefont {R.}~\bibnamefont {Augusiak}},
  \bibinfo {author} {\bibfnamefont {A.}~\bibnamefont {Sainz}}, \bibinfo
  {author} {\bibfnamefont {B.}~\bibnamefont {L{\"u}cke}}, \bibinfo {author}
  {\bibfnamefont {C.}~\bibnamefont {Klempt}}, \bibinfo {author} {\bibfnamefont
  {M.}~\bibnamefont {Lewenstein}},\ and\ \bibinfo {author} {\bibfnamefont
  {A.}~\bibnamefont {Ac{\'i}n}},\ }\href
  {https://doi.org/http://doi.org/10.1016/j.aop.2015.07.021} {\bibfield
  {journal} {\bibinfo  {journal} {Annals of Physics}\ }\textbf {\bibinfo
  {volume} {362}},\ \bibinfo {pages} {370 } (\bibinfo {year}
  {2015})}\BibitemShut {NoStop}%
\bibitem [{\citenamefont {Mekonnen}\ \emph {et~al.}(2025)\citenamefont
  {Mekonnen}, \citenamefont {Galley},\ and\ \citenamefont
  {Mueller}}]{mekonnen2025invariance}%
  \BibitemOpen
  \bibfield  {author} {\bibinfo {author} {\bibfnamefont {M.}~\bibnamefont
  {Mekonnen}}, \bibinfo {author} {\bibfnamefont {T.~D.}\ \bibnamefont
  {Galley}},\ and\ \bibinfo {author} {\bibfnamefont {M.~P.}\ \bibnamefont
  {Mueller}},\ }\href@noop {} {\bibfield  {journal} {\bibinfo  {journal} {arXiv
  preprint arXiv:2502.17576}\ } (\bibinfo {year} {2025})}\BibitemShut {NoStop}%
\bibitem [{\citenamefont {Yaffe}(1982)}]{YeffeRMP}%
  \BibitemOpen
  \bibfield  {author} {\bibinfo {author} {\bibfnamefont {L.~G.}\ \bibnamefont
  {Yaffe}},\ }\href {https://doi.org/10.1103/RevModPhys.54.407} {\bibfield
  {journal} {\bibinfo  {journal} {Rev. Mod. Phys.}\ }\textbf {\bibinfo {volume}
  {54}},\ \bibinfo {pages} {407} (\bibinfo {year} {1982})}\BibitemShut
  {NoStop}%
\bibitem [{\citenamefont {Gnutzmann}\ \emph {et~al.}(1999)\citenamefont
  {Gnutzmann}, \citenamefont {Haake},\ and\ \citenamefont
  {Kus}}]{GnutzmannJPhysA1999}%
  \BibitemOpen
  \bibfield  {author} {\bibinfo {author} {\bibfnamefont {S.}~\bibnamefont
  {Gnutzmann}}, \bibinfo {author} {\bibfnamefont {F.}~\bibnamefont {Haake}},\
  and\ \bibinfo {author} {\bibfnamefont {M.}~\bibnamefont {Kus}},\ }\href
  {https://doi.org/10.1088/0305-4470/33/1/309} {\bibfield  {journal} {\bibinfo
  {journal} {Journal of Physics A: Mathematical and General}\ }\textbf
  {\bibinfo {volume} {33}},\ \bibinfo {pages} {143} (\bibinfo {year}
  {1999})}\BibitemShut {NoStop}%
\bibitem [{\citenamefont {Oganesyan}\ and\ \citenamefont
  {Huse}(2007)}]{oganesyan2007localization}%
  \BibitemOpen
  \bibfield  {author} {\bibinfo {author} {\bibfnamefont {V.}~\bibnamefont
  {Oganesyan}}\ and\ \bibinfo {author} {\bibfnamefont {D.~A.}\ \bibnamefont
  {Huse}},\ }\href@noop {} {\bibfield  {journal} {\bibinfo  {journal} {Physical
  Review B—Condensed Matter and Materials Physics}\ }\textbf {\bibinfo
  {volume} {75}},\ \bibinfo {pages} {155111} (\bibinfo {year}
  {2007})}\BibitemShut {NoStop}%
\bibitem [{\citenamefont {Atas}\ \emph {et~al.}(2013)\citenamefont {Atas},
  \citenamefont {Bogomolny}, \citenamefont {Giraud},\ and\ \citenamefont
  {Roux}}]{atas2013distribution}%
  \BibitemOpen
  \bibfield  {author} {\bibinfo {author} {\bibfnamefont {Y.~Y.}\ \bibnamefont
  {Atas}}, \bibinfo {author} {\bibfnamefont {E.}~\bibnamefont {Bogomolny}},
  \bibinfo {author} {\bibfnamefont {O.}~\bibnamefont {Giraud}},\ and\ \bibinfo
  {author} {\bibfnamefont {G.}~\bibnamefont {Roux}},\ }\href@noop {} {\bibfield
   {journal} {\bibinfo  {journal} {Physical review letters}\ }\textbf {\bibinfo
  {volume} {110}},\ \bibinfo {pages} {084101} (\bibinfo {year}
  {2013})}\BibitemShut {NoStop}%
\bibitem [{\citenamefont {Karampagia}\ \emph {et~al.}(2022)\citenamefont
  {Karampagia}, \citenamefont {Zelevinsky},\ and\ \citenamefont
  {Spitler}}]{karampagia2022ratio}%
  \BibitemOpen
  \bibfield  {author} {\bibinfo {author} {\bibfnamefont {S.}~\bibnamefont
  {Karampagia}}, \bibinfo {author} {\bibfnamefont {V.}~\bibnamefont
  {Zelevinsky}},\ and\ \bibinfo {author} {\bibfnamefont {J.}~\bibnamefont
  {Spitler}},\ }\href@noop {} {\bibfield  {journal} {\bibinfo  {journal}
  {Nuclear Physics A}\ }\textbf {\bibinfo {volume} {1023}},\ \bibinfo {pages}
  {122453} (\bibinfo {year} {2022})}\BibitemShut {NoStop}%
\bibitem [{\citenamefont {M\"uller-Rigat}\ \emph {et~al.}(2024)\citenamefont
  {M\"uller-Rigat}, \citenamefont {Aloy}, \citenamefont {Lewenstein},
  \citenamefont {Fadel},\ and\ \citenamefont
  {Tura}}]{mullerrigat2024threeoutcome}%
  \BibitemOpen
  \bibfield  {author} {\bibinfo {author} {\bibfnamefont {G.}~\bibnamefont
  {M\"uller-Rigat}}, \bibinfo {author} {\bibfnamefont {A.}~\bibnamefont
  {Aloy}}, \bibinfo {author} {\bibfnamefont {M.}~\bibnamefont {Lewenstein}},
  \bibinfo {author} {\bibfnamefont {M.}~\bibnamefont {Fadel}},\ and\ \bibinfo
  {author} {\bibfnamefont {J.}~\bibnamefont {Tura}},\ }\Eprint
  {https://arxiv.org/abs/2406.12823} {arXiv:2406.12823}  (\bibinfo {year}
  {2024})\BibitemShut {NoStop}%
\bibitem [{\citenamefont {Meredith}\ \emph {et~al.}(1988)\citenamefont
  {Meredith}, \citenamefont {Koonin},\ and\ \citenamefont
  {Zirnbauer}}]{MeredithPRA1988}%
  \BibitemOpen
  \bibfield  {author} {\bibinfo {author} {\bibfnamefont {D.~C.}\ \bibnamefont
  {Meredith}}, \bibinfo {author} {\bibfnamefont {S.~E.}\ \bibnamefont
  {Koonin}},\ and\ \bibinfo {author} {\bibfnamefont {M.~R.}\ \bibnamefont
  {Zirnbauer}},\ }\href {https://doi.org/10.1103/physreva.37.3499} {\bibfield
  {journal} {\bibinfo  {journal} {Physical Review A}\ }\textbf {\bibinfo
  {volume} {37}},\ \bibinfo {pages} {3499} (\bibinfo {year}
  {1988})}\BibitemShut {NoStop}%
\bibitem [{\citenamefont {Lipkin}\ \emph {et~al.}(1965)\citenamefont {Lipkin},
  \citenamefont {Meshkov},\ and\ \citenamefont {Glick}}]{LipkinNucPhys1965}%
  \BibitemOpen
  \bibfield  {author} {\bibinfo {author} {\bibfnamefont {H.}~\bibnamefont
  {Lipkin}}, \bibinfo {author} {\bibfnamefont {N.}~\bibnamefont {Meshkov}},\
  and\ \bibinfo {author} {\bibfnamefont {A.}~\bibnamefont {Glick}},\ }\href
  {https://doi.org/10.1016/0029-5582(65)90862-x} {\bibfield  {journal}
  {\bibinfo  {journal} {Nuclear Physics}\ }\textbf {\bibinfo {volume} {62}},\
  \bibinfo {pages} {188} (\bibinfo {year} {1965})}\BibitemShut {NoStop}%
\bibitem [{\citenamefont {Gra\ss{}}\ \emph {et~al.}(2013)\citenamefont
  {Gra\ss{}}, \citenamefont {Juli\'a-D\'{\i}az}, \citenamefont
  {Ku\ifmmode~\acute{s}\else \'{s}\fi{}},\ and\ \citenamefont
  {Lewenstein}}]{GrassPRL2013}%
  \BibitemOpen
  \bibfield  {author} {\bibinfo {author} {\bibfnamefont {T.}~\bibnamefont
  {Gra\ss{}}}, \bibinfo {author} {\bibfnamefont {B.}~\bibnamefont
  {Juli\'a-D\'{\i}az}}, \bibinfo {author} {\bibfnamefont {M.}~\bibnamefont
  {Ku\ifmmode~\acute{s}\else \'{s}\fi{}}},\ and\ \bibinfo {author}
  {\bibfnamefont {M.}~\bibnamefont {Lewenstein}},\ }\href
  {https://doi.org/10.1103/PhysRevLett.111.090404} {\bibfield  {journal}
  {\bibinfo  {journal} {Phys. Rev. Lett.}\ }\textbf {\bibinfo {volume} {111}},\
  \bibinfo {pages} {090404} (\bibinfo {year} {2013})}\BibitemShut {NoStop}%
\bibitem [{\citenamefont {Marconi}\ \emph {et~al.}(2025)\citenamefont
  {Marconi}, \citenamefont {M\"{u}ller-Rigat}, \citenamefont {Romero-Pallejà},
  \citenamefont {Tura},\ and\ \citenamefont {Sanpera}}]{ReviewSymmetrics}%
  \BibitemOpen
  \bibfield  {author} {\bibinfo {author} {\bibfnamefont {C.}~\bibnamefont
  {Marconi}}, \bibinfo {author} {\bibfnamefont {G.}~\bibnamefont
  {M\"{u}ller-Rigat}}, \bibinfo {author} {\bibfnamefont {J.}~\bibnamefont
  {Romero-Pallejà}}, \bibinfo {author} {\bibfnamefont {J.}~\bibnamefont
  {Tura}},\ and\ \bibinfo {author} {\bibfnamefont {A.}~\bibnamefont
  {Sanpera}},\ }\href {https://doi.org/10.48550/ARXIV.2506.10185} {\bibinfo
  {title} {Symmetric quantum states: a review of recent progress}} (\bibinfo
  {year} {2025})\BibitemShut {NoStop}%
\bibitem [{\citenamefont {Trail}\ \emph {et~al.}(2008)\citenamefont {Trail},
  \citenamefont {Madhok},\ and\ \citenamefont
  {Deutsch}}]{trail2008entanglement}%
  \BibitemOpen
  \bibfield  {author} {\bibinfo {author} {\bibfnamefont {C.~M.}\ \bibnamefont
  {Trail}}, \bibinfo {author} {\bibfnamefont {V.}~\bibnamefont {Madhok}},\ and\
  \bibinfo {author} {\bibfnamefont {I.~H.}\ \bibnamefont {Deutsch}},\
  }\href@noop {} {\bibfield  {journal} {\bibinfo  {journal} {Physical Review
  E—Statistical, Nonlinear, and Soft Matter Physics}\ }\textbf {\bibinfo
  {volume} {78}},\ \bibinfo {pages} {046211} (\bibinfo {year}
  {2008})}\BibitemShut {NoStop}%
\bibitem [{\citenamefont {Madhok}\ \emph {et~al.}(2015)\citenamefont {Madhok},
  \citenamefont {Gupta}, \citenamefont {Trottier},\ and\ \citenamefont
  {Ghose}}]{madhok2015signatures}%
  \BibitemOpen
  \bibfield  {author} {\bibinfo {author} {\bibfnamefont {V.}~\bibnamefont
  {Madhok}}, \bibinfo {author} {\bibfnamefont {V.}~\bibnamefont {Gupta}},
  \bibinfo {author} {\bibfnamefont {D.-A.}\ \bibnamefont {Trottier}},\ and\
  \bibinfo {author} {\bibfnamefont {S.}~\bibnamefont {Ghose}},\ }\href@noop {}
  {\bibfield  {journal} {\bibinfo  {journal} {Physical Review E}\ }\textbf
  {\bibinfo {volume} {91}},\ \bibinfo {pages} {032906} (\bibinfo {year}
  {2015})}\BibitemShut {NoStop}%
\bibitem [{\citenamefont {Ramchander}\ and\ \citenamefont
  {Lakshminarayan}(2019)}]{ramchander2019quantum}%
  \BibitemOpen
  \bibfield  {author} {\bibinfo {author} {\bibfnamefont {M.}~\bibnamefont
  {Ramchander}}\ and\ \bibinfo {author} {\bibfnamefont {A.}~\bibnamefont
  {Lakshminarayan}},\ }\href {https://doi.org/10.48550/arXiv.1912.07097}
  {\bibinfo {title} {Quantum chaos and macroscopic realism as no-signaling in
  time}} (\bibinfo {year} {2019})\BibitemShut {NoStop}%
\bibitem [{\citenamefont {{\v{S}}upi{\'{c}}}\ and\ \citenamefont
  {Bowles}(2020)}]{SupicReview}%
  \BibitemOpen
  \bibfield  {author} {\bibinfo {author} {\bibfnamefont {I.}~\bibnamefont
  {{\v{S}}upi{\'{c}}}}\ and\ \bibinfo {author} {\bibfnamefont {J.}~\bibnamefont
  {Bowles}},\ }\href {https://doi.org/10.22331/q-2020-09-30-337} {\bibfield
  {journal} {\bibinfo  {journal} {{Quantum}}\ }\textbf {\bibinfo {volume}
  {4}},\ \bibinfo {pages} {337} (\bibinfo {year} {2020})}\BibitemShut {NoStop}%
\bibitem [{\citenamefont {Vourdas}(2004)}]{AVourdas_2004}%
  \BibitemOpen
  \bibfield  {author} {\bibinfo {author} {\bibfnamefont {A.}~\bibnamefont
  {Vourdas}},\ }\href {https://doi.org/10.1088/0034-4885/67/3/R03} {\bibfield
  {journal} {\bibinfo  {journal} {Reports on Progress in Physics}\ }\textbf
  {\bibinfo {volume} {67}},\ \bibinfo {pages} {267} (\bibinfo {year}
  {2004})}\BibitemShut {NoStop}%
\bibitem [{\citenamefont {Shurtleff}(2023)}]{shurtleff2023formulas}%
  \BibitemOpen
  \bibfield  {author} {\bibinfo {author} {\bibfnamefont {R.}~\bibnamefont
  {Shurtleff}},\ }\href@noop {} {\bibinfo {title} {Formulas for su(3)
  matrices}} (\bibinfo {year} {2023}),\ \Eprint
  {https://arxiv.org/abs/0908.3864} {arXiv:0908.3864 [math-ph]} \BibitemShut
  {NoStop}%
\bibitem [{\citenamefont {Brody}(1973)}]{brody1973statistical}%
  \BibitemOpen
  \bibfield  {author} {\bibinfo {author} {\bibfnamefont {T.~A.}\ \bibnamefont
  {Brody}},\ }\href {https://doi.org/10.1007/BF02727859} {\bibfield  {journal}
  {\bibinfo  {journal} {Lettere al Nuovo Cimento (1971-1985)}\ }\textbf
  {\bibinfo {volume} {7}},\ \bibinfo {pages} {482} (\bibinfo {year}
  {1973})}\BibitemShut {NoStop}%
\bibitem [{\citenamefont {Bittner}\ \emph {et~al.}(2001)\citenamefont
  {Bittner}, \citenamefont {Markum},\ and\ \citenamefont
  {Pullirsch}}]{bittner2001quantum}%
  \BibitemOpen
  \bibfield  {author} {\bibinfo {author} {\bibfnamefont {E.}~\bibnamefont
  {Bittner}}, \bibinfo {author} {\bibfnamefont {H.}~\bibnamefont {Markum}},\
  and\ \bibinfo {author} {\bibfnamefont {R.}~\bibnamefont {Pullirsch}},\
  }\Eprint {https://arxiv.org/abs/hep-lat/0110222} {arXiv:hep-lat/0110222
  [hep-lat]}  (\bibinfo {year} {2001})\BibitemShut {NoStop}%
\end{thebibliography}%

\clearpage
\newpage

\begin{widetext}

\section{Methods and Supplementary Material}

\subsection{Classical bound for the Bell inequality~(1) in the main text}
\label{app:classicalBound}

Here we provide a proof showing that the Bell inequality $B$ introduced in the main text has classical bound $\beta_c = 0$ for any number of parties $n$. That is, we want to show that
\begin{eqnarray}
B&=& (\mathcal{P}_{0|0}+\mathcal{P}_{0|1}+ \mathcal{P}_{1|0}+\mathcal{P}_{1|1}) + 
(\mathcal{P}_{00|00}+\mathcal{P}_{00|11}+\mathcal{P}_{11|00}+\mathcal{P}_{11|11}) 
- 2(\mathcal{P}_{01|01}+\mathcal{P}_{01|10}) \geq 0,
\label{eq:PIBIappendix}
\end{eqnarray}
where recall that $\mathcal{P}_{a|x}=\sum_{i\in [n]}p_i(a|x)$ is the collective one-body conditional probability, with $p_i(a|x)$ denoting the probability that subsystem $i$ yields outcome $a$ given measurement setting $x$. Similarly, $\mathcal{P}_{ab|xy}=\sum_{i\neq j \in [n]}p_{ij}(ab|xy)$ represents the collective two-body conditional probability summing over all possible pairs $i\neq j\in [n]$.

First, in order to account for all local hidden variables, we want to come up with a parametrization to describe the conditional probabilities in terms of Local Deterministic Strategies (LDS). Because our Bell inequality is permutationally invariant, many Bell inequality coefficients take the same values at many LDSs, which leads to redundancies. Hence, instead of considering all the $d^{mn}$ possibilities of the general case ($3^{2n}$ for our case with $3$ outcomes an $2$ measurements), we propose the following (much more efficient) parametrization: Suppose that at each run all the outcomes for each possible measurement and party are predetermined. Then, let $c_{a,a'}$ be the total number of parties that have predetermined the pair of outcomes $a,a'\in \{0,1,2\}$ for the two possible measurement settings $x\in\{0,1\}$ respectively. It follows by definition that $c_{a,a'}\geq 0$ and $\sum_{a,a'}c_{a,a'}=n$. Therefore, following this parametrization, the symmetrized one-body conditional probabilities $\mathcal{P}_{a|x}$ under a given LDS can be expressed as:
\begin{align}
\mathcal{P}_{a|x}:=\sum\limits_{i=1}^{n}p_i(a|x)\LDSeq\left\lbrace\begin{array}{cc}
    c_{a,0}+c_{a,1}+c_{a,2} & \textrm{for }x=0 \\
    c_{0,a}+c_{1,a}+c_{2,a} & \textrm{for }x=1
\end{array}\right. .
\label{eq:oneProb}
\end{align}
On the other hand, the symmetric two-body conditional probabilities $\mathcal{P}_{ab|xy}$ factorize under a given LDS as:
\begin{equation} 
\begin{split}
 \mathcal{P}_{ab|xy} :=& \sum\limits_{i\neq j}p_{ij}(ab|xy) \LDSeq \sum\limits_{i\neq j}p_i(a|x)p_j(b|y) \\
=& \underbrace{\sum\limits_{i\neq j}p_{i}(a|x)p_j(b|y)+\sum\limits_i p_i(a|x)p_i(b|y)}_{\mathcal{P}_{a|x}\cdot \mathcal{P}_{b|y}} -\underbrace{\sum\limits_i p_i(a|x)p_i(b|y)}_{:=\mathcal{Q}_{ab|xy}} \;, \label{eq:twoProb}
\end{split}
\end{equation}
where we have defined
\begin{align}
\mathcal{Q}_{ab,xy}:=
\left\{\begin{array}{l}\mathcal{P}_{a|x}\\
0 \\
c_{a,b}\\ c_{b,a}
\end{array}\right.
\begin{array}{l}\text{if }a=b,x=y\\\text{if }a\neq b,x=y\\ \text{if }x=0, y=1\\ \text{if }x=1, y=0 \end{array} \;.
\label{eq:Q}
\end{align} 
Note that one can neglect one of the outcomes without loss of generality by means of the NS principle,
\begin{equation}\label{eq:NS}
 P(a_1, \ldots, \hat{a}_i, \ldots, a_{n}| x_1, \ldots, \hat{x}_i, \ldots, x_{n}) \equiv  \sum\limits_{a_i \in \{0,1,2\}}P(a_1,\ldots, a_n|x_1,\ldots, x_n) \;, \nonumber
\end{equation}
where $\hat{\cdot}$ denotes the absence of that coordinate and the $\equiv$ symbol means that the LHS of \cref{eq:NS} is well-defined; \textit{i.e.}, it does not depend on the value of $x_i$. Hence, for instance we can take \cref{eq:oneProb,eq:twoProb,eq:Q} with $a,b\in\{0,1\}$. Notice also that it is straightforward to generalize our parametrization to any number of outcomes $d$.

Finally, in Tab.~\ref{table:LDS} we express the one-body terms and the factorized two-body terms as a function of the quantities $c_{a,a'}$. Therefore, all the possible local-realist correlations in a $(n,2,3)$ Bell-type experiments can be described in terms of the relations in \Cref{table:LDS} and shared randomness. Moreover, the local polytope for an $(n,2,3)$ permutationally invariant Bell scenario characterized by one- and two-body correlators is formed by the convex hull of all the configurations satisfying the relations in \Cref{table:LDS}. 

\begin{table*}[h!]
\centering
\begin{tabular}{l@{\hskip 0.05in}|@{\hskip 0.05in}l@{\hskip 0.05in}|@{\hskip 0.05in}l@{\hskip 0.05in}|@{\hskip 0.05in}l@{\hskip 0.05in}|@{\hskip 0.05in}l}
$\mathcal{P}_{0|0}\LDSeq c_{0,0}+c_{0,1}+c_{0,2}$ & $\mathcal{P}_{00|00}\LDSeq\mathcal{P}_{0|0}^2-\mathcal{P}_{0|0}$ & $\mathcal{P}_{00|01}\LDSeq\mathcal{P}_{0|0}\mathcal{P}_{0|1}-c_{0,0}$ & $\mathcal{P}_{00|10}\LDSeq\mathcal{P}_{00|01}$ & $\mathcal{P}_{00|11}\LDSeq\mathcal{P}_{0|1}^2-\mathcal{P}_{0|1}$\\
$\mathcal{P}_{1|0}\LDSeq c_{1,0}+c_{1,1}+c_{1,2}$ & $\mathcal{P}_{01|00}\LDSeq\mathcal{P}_{0|0}\mathcal{P}_{1|0}$ & $\mathcal{P}_{01|01}\LDSeq\mathcal{P}_{0|0}\mathcal{P}_{1|1}-c_{0,1}$ & $\mathcal{P}_{01|10}\LDSeq\mathcal{P}_{10|01}$ & $\mathcal{P}_{01|11}\LDSeq\mathcal{P}_{0|1}\mathcal{P}_{1|1}$\\
$\mathcal{P}_{0|1}\LDSeq c_{0,0}+c_{1,0}+c_{2,0}$ & $\mathcal{P}_{10|00}\LDSeq\mathcal{P}_{01|00}$ & $\mathcal{P}_{10|01}\LDSeq\mathcal{P}_{1|0}\mathcal{P}_{0|1}-c_{1,0}$ & $\mathcal{P}_{10|10}\LDSeq\mathcal{P}_{01|01}$ & $\mathcal{P}_{10|11}\LDSeq\mathcal{P}_{01|11}$ \\
$\mathcal{P}_{1|1}\LDSeq c_{0,1}+c_{1,1}+c_{2,1}$ & $\mathcal{P}_{11|00}\LDSeq\mathcal{P}_{1|0}^2-\mathcal{P}_{1|0}$ &  $\mathcal{P}_{11|01}\LDSeq\mathcal{P}_{1|0}\mathcal{P}_{1|1}-c_{1,1}$ & $\mathcal{P}_{11|10}\LDSeq\mathcal{P}_{11|01}$ & $\mathcal{P}_{11|11}\LDSeq\mathcal{P}_{1|1}^2-\mathcal{P}_{1|1}$
\end{tabular}
\caption{Resulting one- and two-body conditional probabilities under an LDS.}\label{table:LDS}
\end{table*}

Now that we have a parametrization to incorporate the LDSs into the conditional probabilities, we are ready to substitute the corresponding conditional probabilities in the Bell inequality. After rearranging the terms, one ends up with the following polynomial:
\begin{eqnarray}
B&=&(c_{00}+c_{02})^2 + (c_{00}+c_{20})^2 + (c_{11}+c_{12})^2 + (c_{11}+c_{21})^2 \nonumber\\
&&+ (c_{00}-c_{12})^2 + (c_{00}-c_{21})^2 + (c_{11}-c_{02})^2 + (c_{11}-c_{20})^2 \nonumber\\
&& +  2(c_{10} + c_{01}) - 2(c_{00} + c_{11})^2 - (c_{12} + c_{20})^2 - (c_{02} + c_{21})^2,
\label{eq:poly}
\end{eqnarray}
where $c_{i,j}\geq 0$ for all $i,j\in \{0,1,2\}$ and they fulfill the constraint $\sum\limits_{0\leq i,j<3}c_{ij}=n$ with $n$ the total number of parties. Notice that the term $c_{22}$ does not appear in the expression, thus we can set any $0\leq c_{22}\leq n$ without contributing in the classical bound. Thus it is trivial to see that there exists at least one strategy leading to $B=0$, \textit{i.e.} setting $c_{22}=n$. Consequently, we just have to prove that $B$ cannot take negative values. \\

\paragraph*{Proof that $B\geq 0$:} We are interested in the minimal value that \eqref{eq:poly} can achieve. Since the terms $2(c_{10}+c_{01})$ will always add a positive or zero contribution, we can set them to $c_{10}=c_{01}=0$ without loss of generality to find the minimal value of $B$. Therefore we simplify the problem to look at the minimal value of:

\begin{eqnarray}
\tilde{B}&=&(c_{00}+c_{02})^2 + (c_{00}+c_{20})^2 + (c_{11}+c_{12})^2 + (c_{11}+c_{21})^2 \nonumber\\
&&+ (c_{00}-c_{12})^2 + (c_{00}-c_{21})^2 + (c_{11}-c_{02})^2 + (c_{11}-c_{20})^2 \nonumber\\
&&  - 2(c_{00} + c_{11})^2 - (c_{12} + c_{20})^2 - (c_{02} + c_{21})^2.
\end{eqnarray}

After expanding and rearranging the terms we reach the following equivalent polynomial:
\begin{eqnarray}
\tilde{B}&=&2 \bigg[ c_{00}^2+c_{11}^2+\frac{c_{02}^2+c_{20}^2}{2}+\frac{c_{12}^2+c_{21}^2}{2} \nonumber\\
&& +c_{00}(c_{02}+c_{20})-c_{11}(c_{02}+c_{20}) +c_{11}(c_{12}+c_{21})-c_{00}(c_{12}+c_{21}) \nonumber\\
&& -c_{02}c_{21}-c_{12}c_{20}-2c_{00}c_{11} \bigg].
\label{eq:poly2}
\end{eqnarray}

Then, the condition for \eqref{eq:poly2} to take negative values corresponds to the following inequality
\begin{equation}
c_{11}(c_{02}+c_{20})+c_{00}(c_{12}+c_{21})+c_{02}c_{21}+c_{12}c_{20}+2c_{00}c_{11} > c_{00}^2+c_{11}^2+\frac{c_{02}^2+c_{20}^2}{2}+\frac{c_{12}^2+c_{21}^2}{2}+c_{00}(c_{02}+c_{20})+c_{11}(c_{12}+c_{21}) \;,
\end{equation}

which can be rearranged as:

\begin{equation}
(c_{00}-c_{11})(c_{12}+c_{21}-c_{02}-c_{20}) > (c_{00}-c_{11})^2+\frac{(c_{02}-c_{21})^2}{2}+\frac{(c_{12}-c_{20})^2}{2} \;.
\label{eq:wantContradiction}
\end{equation}

Our goal is to find that such condition leads to a contradiction for all cases in order to show that $I$ cannot take a negative value. 

First it is convenient to define de variables $x:=c_{00}-c_{11}, y:=c_{12}-c_{20}, z:=c_{02}-c_{21}$, so that the condition gets expressed as:

\begin{equation}
x(y-z) - \left(x^2+\frac{y^2}{2}+\frac{z^2}{2} \right) > 0.
\label{eq:wantContradiction2}
\end{equation}

Take now $f(x,y,z)= x(y-z) - (x^2+\frac{y^2}{2}+\frac{z^2}{2})$ in order to find its critical points $\nabla f(x,y,z)=(-2x+y-z,x-y,-z-x)$, $\nabla f(x^*,y^*,z^*)=0 \Rightarrow x^*=y^*,z^*=-x^*$, where $f(x^*,y^*,z^*)=0$. Next, by looking at its Hessian matrix $\bm{H}\left(f(x,y,z)\right)$, where $\left(\bm{H}\left(f(x,y,z)\right)\right)_{ij}=\frac{\partial^2 f}{\partial x_i\partial x_j}$, one sees that the resulting Hessian matrix has eigenvalues $\{-3,-1,0\}$ and therefore it is negative semidefinite. Thus, the critical point corresponds to the maximum.

We conclude that \eqref{eq:wantContradiction2} leads to a contradiction for all values of $c_{ij}$ and, consequently, $I$ cannot take negative values. Finally, since the argument is independent of $n$ and we have seen that $I=0$ is a valid local deterministic strategy, it follows that the classical bound is $\beta_c=0$ for all $n$. \\

\hfill\ensuremath{\square}

\subsection{Measurement Parametrization for Qutrits}
\label{app:OptimizationParametrization}

For two-level systems (\ie qubits), projective measurements can be naturally parameterized using Pauli operators and linear combinations thereof, which maintain unitarity and allow for smooth interpolation between different measurement bases. However, extending this approach to higher-dimensional systems such as qutrits is not straightforward. The main difficulty lies in constructing continuous families of unitary operators that preserve the spectral properties required for the Bell scenario. 

For example, the straightforward generalization of Pauli matrices $\sigma_x$ and $\sigma_z$ to a three-level system are the Heisenberg-Weyl observables $X$ and $Z$, where $X$ shifts the standard basis as $\ket{0} \mapsto \ket{1} \mapsto \ket{2}$ and $Z$ applies a third root of unity $\ket{\alpha} \mapsto \zeta^{\alpha} \ket{\alpha}$, with $\zeta = e^{-2\pi \mathbbm{i}/3}$. More general, for any $d$-level system, 
\[X\ket{\alpha}=\ket{\alpha+1 \mod d}, \quad Z\ket{\alpha}=\zeta^{\alpha}\ket{\alpha}
,\]
where both operators satisfy $X^d = Z^d = \mathbb{I}$ \cite{AVourdas_2004}. However, note that, in general, for $d>2$ a unit vector $\boldsymbol{u} = (u_x, u_z)$ will not preserve unitarity of $u_x X + u_z Z$.

To address this, we propose a strategy that uses a set of $M$ unitaries $\{U_0, \ldots, U_{M-1}\}$ sharing a fixed spectrum ordered by the complex phase argument of the $d$-roots of unity $\{1, \zeta, \ldots, \zeta^{d-1}\}$, where $d=3$ in the qutrits case. For concreteness, in our qutrit implementation, one of the choices that heuristically yielded a good compromise between efficiency and accuracy is the following set of $8$ unitaries $\{X,Z,X^2,XZ,ZX,XZ^2,Z^2X,X^2Z^2\}$. 
Then, instead of directly parametrizing by combining these unitaries, we shift the parametrization to interpolate through Hermitian matrices $g_k$, from which we construct unitary matrices that retain said spectrum. In particular, we define
\begin{equation}\label{eq:parametrization}
    g(\boldsymbol{\theta}) := g_0 + \sum\limits_{\ell=1}^M\theta_\ell (g_\ell - g_0),
\end{equation}
where $\boldsymbol{\theta}\in\mathbb{R}^M$ is a vector of $M$ real parameters that controls the interpolation between the Hermitian matrices. Then, we obtain the desired unitary parametrization by defining
\begin{equation}\label{eq:unitaryparametrized}
    U(\boldsymbol{\theta}) := e^{\iu g(\boldsymbol{\theta})}D e^{-\iu g(\boldsymbol{\theta})},
\end{equation}
where $D=\textrm{diag}\left(1,\zeta,\ldots,\zeta^{d-1}\right)$ is a diagonal matrix composed of the $d$-th roots of unity ensuring that the resulting unitary retains the same spectrum as the original set of unitaries. Therefore, by varying $g_\ell$ using the parametrization in \cref{eq:parametrization}, we have obtained an interpolating function $ U(\boldsymbol{\theta})$ in \cref{eq:unitaryparametrized}, which allows us to represent the Bell operator while implementing typical non-convex optimization techniques such as numerical see-saw and stochastic gradient descent.

Finally, through an inverse Fourier transform, each parametrized measurement setting $\boldsymbol{\theta}_x$ defines a projective measurement  
\[P_{a|x}(\bm{\theta_x})=\left(\zeta^{0\cdot a}U(\bm{\theta}_x)^3+\zeta^{1\cdot a}U(\bm{\theta}_x)^2+\zeta^{2\cdot a}U(\bm{\theta}_x)\right)/3,\] 
where $a\in \{0,1,2\}$ labels the measurement outcomes and $x\in\{0,1\}$ the measurement settings. This construction naturally extends to arbitrary $d$-level systems and enables consistent Bell operator definitions that respect the symmetries of the scenario while allowing for continuous variation in measurement parameters. For example, we can now parametrize the first term of  as 
\[\mathcal{P}_{0|0}= \sum\limits_{i=0}^n \braket{P_{0|0}^{(i)}}= \frac{1}{3}\sum\limits_{i=1}^n \braket{ U(\bm{\theta}_0^{(i)})^3+U(\bm{\theta}_0^{(i)})^2+U(\bm{\theta}_0^{(i)}) }, \]
while for the two-body terms, for instance $\mathcal{P}_{01|01}$, one has
\[
\mathcal{P}_{01|01}=\sum\limits_{i\neq j}\braket{P_{0|0}\otimes P_{1|1}}=\frac{1}{9}\sum\limits_{i\neq j}^3\braket{\left(U(\bm{\theta}_0^{(i)})^3+U(\bm{\theta}_0^{(i)})^2+U(\bm{\theta}_0^{(i)}) \right) \otimes \left(U(\bm{\theta}_1^{(j)})^3+\zeta U(\bm{\theta}_1^{(j)})^2+ \zeta^2 U(\bm{\theta}_1^{(j)}) \right) }.
\]
Recall that, in practice, we take the simplification that each party sets the same measurement settings $\bm{\theta}_x^{(i)}=\bm{\theta}_x^{(j)}=\bm{\theta}_x$.

\subsubsection{Measurement optimization restricted to some irreducible representation of SU$(3)$}
\label{app:OptimizationSU3irreps}

In the task of optimizing the measurement settings for a specific irrep, one encounters the problem of representing the symmetrized version of a one-body observable $A$ in that specific irrep. Normally, general formulas for an irrep $(p,q)$ exist only for specific matrices, so one has to do the appropriate change of coordinates to represent an arbitrary $A$ in the irrep $(p,q)$. Here we outline this method.
Following the notation of Ref.~\cite{shurtleff2023formulas}, we start by fixing the eight basis matrices with the identity element $\{\mathbb{I}, T_+, T_-,T^3, V_+, V_-, U_+, U_-, U^3\}^{(p,q)}$ of the finite dimensional $(p,q)-$irrep of SU$(3)$. In Ref.~\cite{shurtleff2023formulas} one can find a possible representation of these for any $(p,q)$. Then, define $G$ as the Gram matrix formed from this basis for the local irrep $(p,q)=(1,0)$; \textit{i.e.,} $G_{i,j} = \braket{v_i, v_j}$, for $v_i \in \{\mathbb{I}, T_+, T_-,T^3, V_+, V_-, U_+, U_-, U^3\}^{(1,0)}$. Next, define a column-vector $\boldsymbol{b} = (\braket{v_1,A},\braket{v_2,A},\ldots,\braket{v_9,A})^T$ consisting of the inner products of this basis for some one-party observable $A$ parametrized as described in the text. Finally, solve the system of linear equations $G\boldsymbol{x}=\boldsymbol{b}$ in order to find $\boldsymbol{x}$. We are now ready to represent an observable $A$ in any chosen irrep $(p,q)$ through the following expression
\begin{equation}
    A^{(p,q)} =  nx_1 \mathbb{I}^{(p,q)} + \sum\limits_{i=2}^9 x_i v_i^{(p,q)}  \;,
\end{equation}
where the first term has been taken out of the sum to go together with the extra identity elements that need to be added in order to guarantee the proper normalization.

Now, if we choose the one-party observable $A$ to be a one-body projective unitary qudit operator (for example, $A=\mathcal{P}_{0|0}$), we can use the approach introduced in the main text, which allows us to perform typical optimization techniques restricted to the chosen irrep $(p,q)$ subsector. Finally, the two-body terms take the form
\begin{equation}
   \mathcal{P}_{ab|xy}^{(p,q)} = \mathcal{P}_{a|x}^{(p,q)}\mathcal{P}_{b|y}^{(p,q)} - (\mathcal{P}_{a|x}\mathcal{P}_{b|y})^{(p,q)} \;.
\end{equation}

\begin{figure}
\centering
\includegraphics[width=0.495\linewidth]{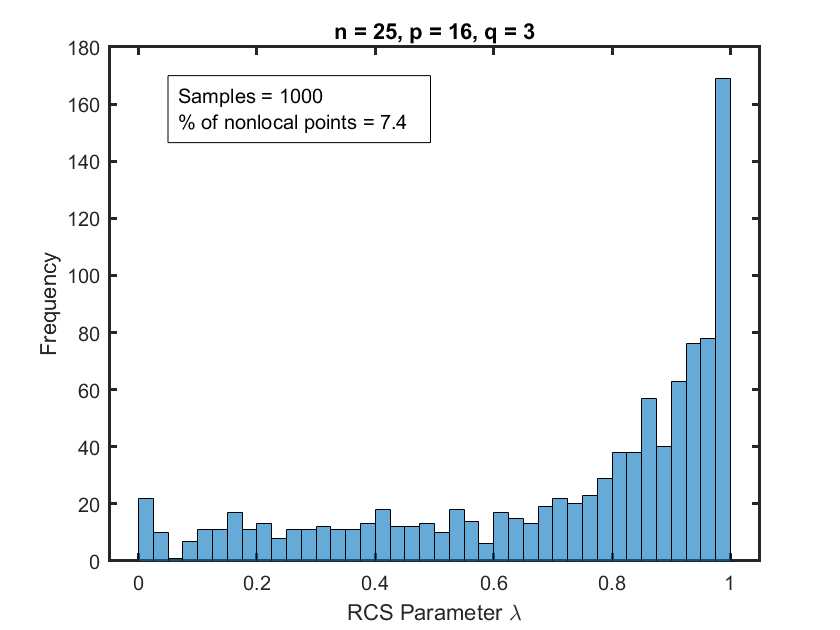}
\hfill
\includegraphics[width=0.495\linewidth]{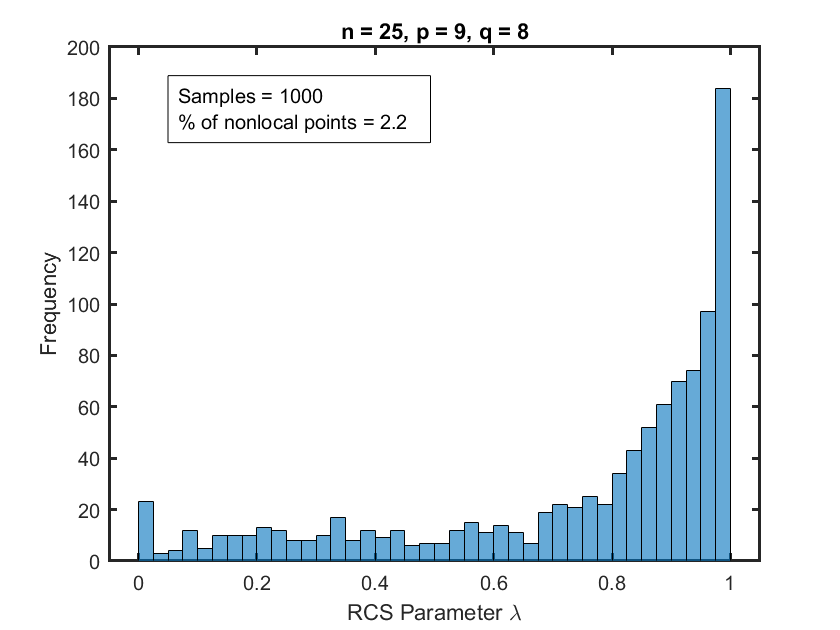}
\\
\includegraphics[width=0.495\linewidth]{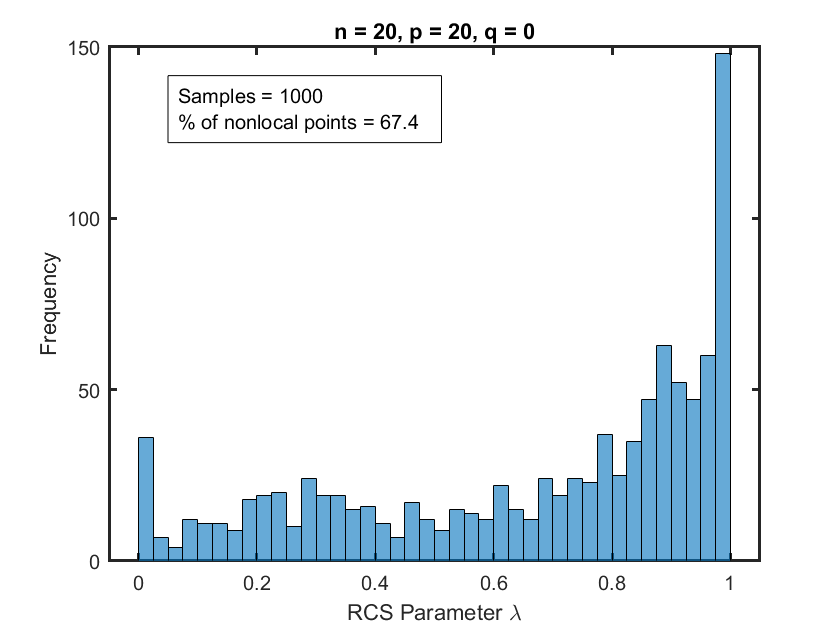}
\hfill
\includegraphics[width=0.495\linewidth]{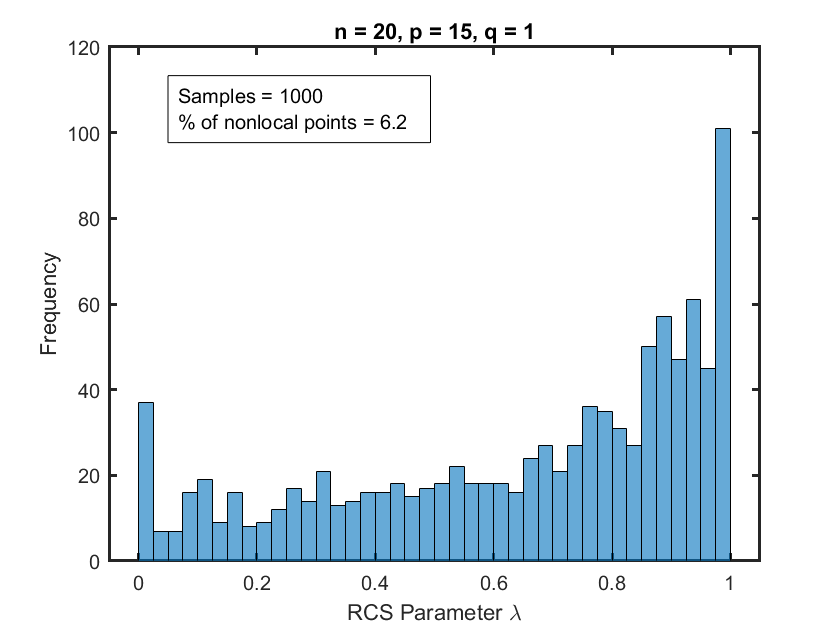}
\caption{More histograms for different irreps with $n=25$ (top) and $n=20$ (bottom), using $10^3$ random projector samples per case, to illustrate that the trend observed in Fig.~\ref{fig3} is generic across various irreps.
That is, all cases show a clear departure from the Poisson distribution, consistent with chaotic behavior in the Bell operator induced by random measurements. 
}
\label{fig:otherRandProj}
\end{figure}

\subsection{Random measurements generation}

The random measurements have been generated by sampling independent $3\times 3$ unitary matrices $U_0$ and $U_1$, associated with the two measurement settings $x\in\{0,1\}$, uniformly according to the Haar measure. For each $U_x$, we define $\tilde U_x = U_x / \det(U_x)^{1/3}$ to enforce unit determinant, and obtain the projective measurement operators $P_{a|x} = \mathbf v_{a|x}\mathbf v_{a|x}^\dagger$, where $\{\mathbf v_{a|x}\}_{a=0}^2$ are the eigenvectors of $\tilde U_x$ from its spectral decomposition $\tilde U_x = V_x D_x V_x^\dagger$.

\subsection{Using the Nearest-Neighbor Spacing Distribution (NNSD)}
\label{app:NNSD}
Before adopting the ratio of consecutive level spacings (RCS) as our main chaos indicator, we initially performed our analysis using the traditional nearest-neighbor spacing distribution (NNSD). While NNSD has been extensively used in the literature to probe quantum chaos~\cite{Haake2018}, it requires spectrum unfolding (see below) and is sensitive to binning choices and other parameters, which can introduce ambiguities. Despite these drawbacks, our findings using NNSD and the Brody distribution interpolation are consistent with the conclusions based on RCS presented in the main text.

In this supplementary section, for completion and comparison, we provide said NNSD results. Rather than repeating the full analysis, we summarize the key steps and highlight where the NNSD confirms the insights obtained through the RCS method. The methodology is the same as the one presented in the main text, except that in the last step we compute the NNSD of the resulting Bell operator (instead of the RCS) and fit it to the Brody distribution Eq.~\eqref{eq:brodyDistribution} to extract the parameter $\omega$ as we explain in what proceeds.

\vspace{2mm}
\textbf{Fitting to the Brody Distribution.—}
If the NNSD follows a Poisson distribution, then the NNSD typically indicates ``level attraction'' in the spectrum, signalling an integrable system in the classical limit. If, on the contrary, the NNSD follows a Wigner-Dyson distribution, then the NNSD typically illustrates ``level repulsion'', which signals a chaotic system in the classical limit (equivalently described by random matrix theory). 
For this reason, it is convenient to fit the NNSD computed numerically with the so-called Brody distribution \cite{brody1973statistical,bittner2001quantum}, which gives a function interpolating between these two limit cases. In particular, for random matrices sampled from the Gaussian orthogonal ensemble, after unfolding the spectrum within each irreducible representation (irrep) of SU(3), the Brody distribution fit of a level spacing distribution $P(s)$ is~\cite{brody1973statistical,bittner2001quantum},
\begin{equation}
    P(s,\omega)=A (\omega+1)s^{\omega}\exp\left(-A s^{\omega+1}\right),
    \label{eq:brodyDistribution}
\end{equation}
where $A=\left(\Gamma\left(\frac{\omega+2}{\omega+1}\right)\right)^{\omega+1}$ with $\Gamma$ denoting the gamma function, and $\omega \in [0,1]$ is the Brody parameter interpolating between the Poisson ($\omega = 0$) and Wigner-Dyson ($\omega = 1$) statistics.

\vspace{2mm}
\textbf{PIBI irreducible representations and NNSD.--}  In  \Cref{fig:RCS_and_NNSD_n25} we show the RCS and NNSD results for $n=25$, along with the corresponding interpolating parameters $\lambda$ and $\omega$ displayed under each irrep for comparison. For better readability, these values are also listed in \Cref{tab:rcs_nnsd_n25}. As discussed in the main text, \cref{fig:RCS_and_NNSD_n25} indicates whether the RCS and NNSD of a given irrep are characteristic of a Poissonian distribution (blue, $\lambda=0$ \& $\omega = 0$, signalling integrability) or a Wigner-Dyson distribution (orange, $\lambda>0$ \& $\omega > 0$, signalling chaos). 

We exclude irreps that do not exhibit nonlocality detection (\textit{i.e.,} those for which $\langle\mathcal{B}\rangle\geq 0$ for all measurement choices) from our analysis, since such cases allow for trivial strategies saturating the classical bound $\langle\mathcal{B}\rangle = 0$ independently of the irrep and do not provide a meaningful setting to probe the relation between nonlocality and spectral statistics. In particular, one such strategy consists in having the same measurement settings in both input possibilities (\textit{i.e., }$\bm{\theta}_0=\bm{\theta}_1$).

One observes that, even though the NNSD results are in principle less robust and dependent on the unfolding procedure, the same behavior is observed across both analysis. That is, when restricting to the optimal measurements to obtain the lowest value $\langle\mathcal{B}\rangle$, irreps that exhibit nonlocality (\textit{i.e., $\langle\mathcal{B}\rangle < 0$}) generically display a Poisson RCS/NNSD. Conversely, irreps where no detection of nonlocality is observed generically display a Wigner-Dyson RCS/NNSD. We observe this behavior for different number of parties $n$, starting from $n=8$ qutrits (when the PIBI we use starts detecting nonlocality) up to $n=32$ (beyond which a numerical analysis becomes computationally expensive). For example, in \Cref{fig:RCS_and_NNSD_n20} and \Cref{tab:rcs_nnsd_n20} we also present the $n=20$ qutrits case for comparison.
This observation leads us to believe that for the Bell operator of the PIBI~\eqref{eq:PIBIappendix} with measurements yielding maximal nonlocality detectability, irreps $(p,q)$ have an RCS fitted with $\lambda = 0$ as $n$ goes to infinity, signalling integrability. As we have seen in the main text, this integrability is likely explained by the additional parity symmetry that occurs around such optimal measurements.
Note that in both cases there are some irreps apparently contradicting this conjecture (see the orange points in \Cref{fig:RCS_and_NNSD_n25,fig:RCS_and_NNSD_n20}). However, we attribute such occurrences to finite size effects, since for $n=25$ or $n=20$ the scarce number of energy levels (after removing redundancies due to the permutation invariance) results in coarse-grained RCS/NNSDs.

\begin{figure}[th]
    \centering
    \includegraphics[width=0.75\linewidth]{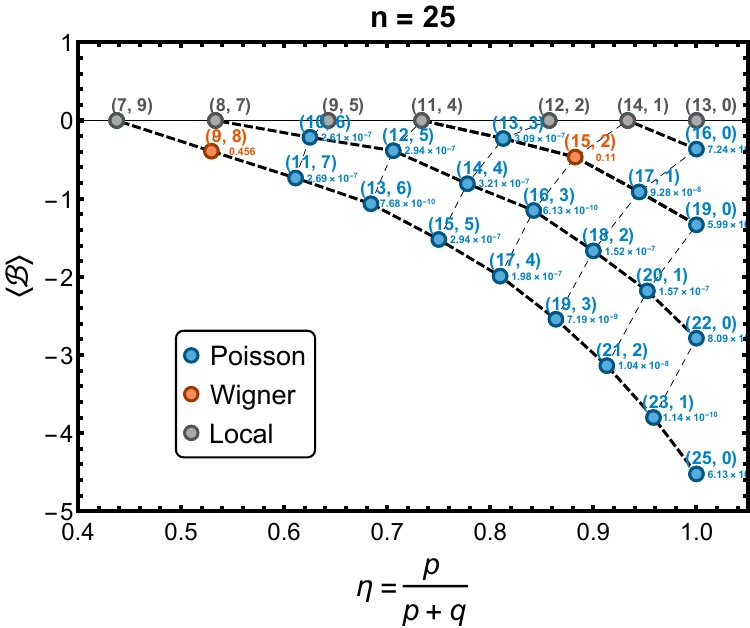}

    \vspace{2em} 

    \includegraphics[width=0.75\linewidth]{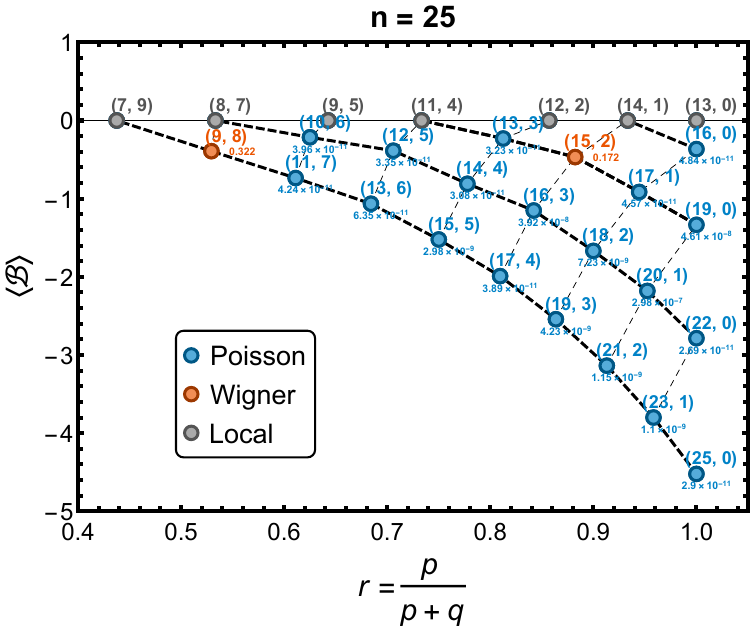}
    
    \caption{(a) Same as the main text Fig.~2, now including the interpolating parameter $\lambda$ fitted to the RCS histograms. 
    (b) Nearest-neighbor level spacing (NNSD) analog of (a), showing the Brody interpolation parameter $\omega$ as small labels. 
    All numerical values are listed in \Cref{tab:rcs_nnsd_n25}.}
    \label{fig:RCS_and_NNSD_n25}
\end{figure}

\begin{figure}[th]
    \centering
    \includegraphics[width=0.75\linewidth]{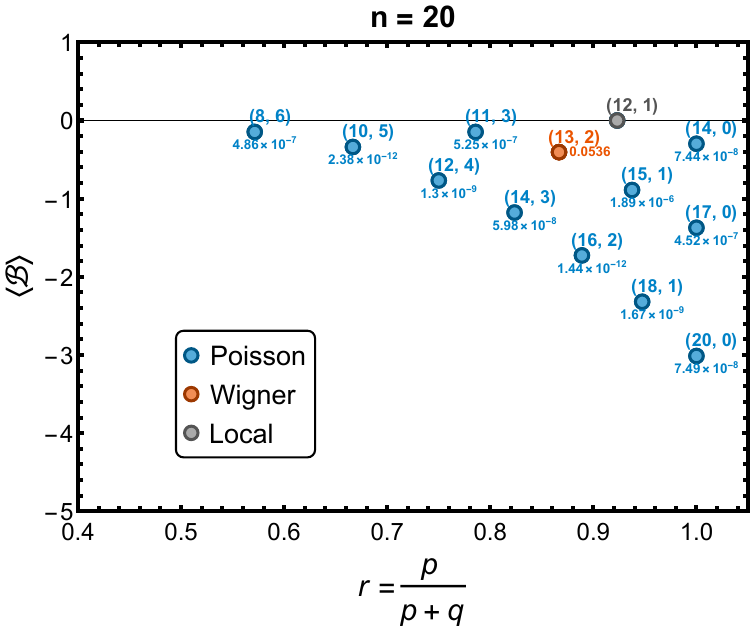}
    
    \vspace{2em} 

    \includegraphics[width=0.75\linewidth]{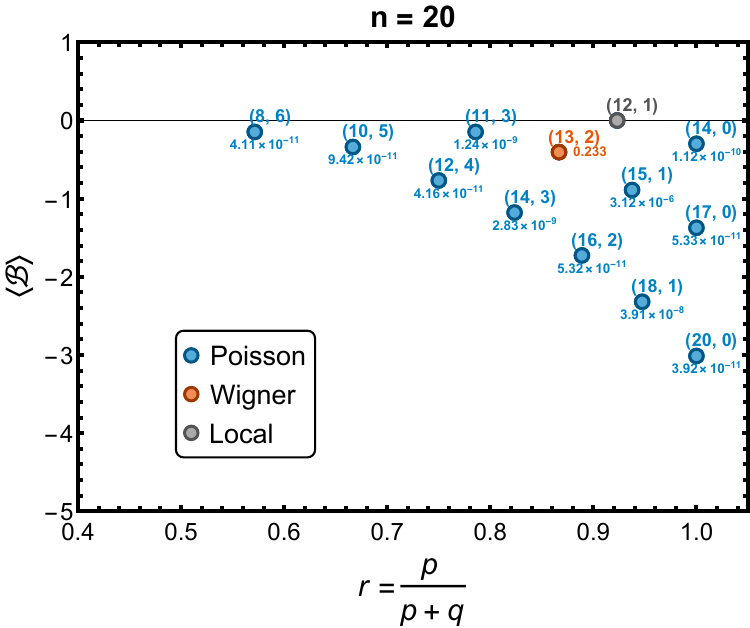}
    
    \caption{(a) Same as \cref{fig:RCS_and_NNSD_n25} but for $n=20$ qutrits. (b) Nearest-neighbor level spacing (NNSD) analog of (a), showing the Brody interpolation parameter $\omega$. The data shown numerically in both plots is summarized in \Cref{tab:rcs_nnsd_n20}.}
    \label{fig:RCS_and_NNSD_n20}
\end{figure}

\begin{table}[]
    \centering
    \begin{tabular}{c|c|c|c}
       \textbf{Irrep} & \textbf{RCS} $\boldsymbol{\lambda}$ & \textbf{NNSD (Brody)} $\boldsymbol{\omega}$ & $\langle \mathcal{B} \rangle$ \\
       \hline
        (9,8) & $4.559 \times 10^{-1}$ & $3.217 \times 10^{-1}$ & $-0.393$ \\
        (14,1) & $2.397 \times 10^{-8}$ & $3.754 \times 10^{-12}$ & $6.939 \times 10^{-10}$ \\
        (13,3) & $3.092 \times 10^{-7}$ & $3.229 \times 10^{-11}$ & $-0.231$ \\
        (12,5) & $2.941 \times 10^{-7}$ & $3.347 \times 10^{-11}$ & $-0.386$ \\
        (11,7) & $2.692 \times 10^{-7}$ & $4.241 \times 10^{-11}$ & $-0.736$ \\
        (16,0) & $7.240 \times 10^{-8}$ & $4.845 \times 10^{-11}$ & $-0.366$ \\
        (15,2) & $1.098 \times 10^{-1}$ & $1.723 \times 10^{-1}$ & $-0.468$ \\
        (14,4) & $3.208 \times 10^{-7}$ & $3.076 \times 10^{-11}$ & $-0.809$ \\
        (13,6) & $7.680 \times 10^{-10}$ & $6.352 \times 10^{-11}$ & $-1.065$ \\
        (17,1) & $9.281 \times 10^{-8}$ & $4.574 \times 10^{-11}$ & $-0.914$ \\
        (16,3) & $6.129 \times 10^{-10}$ & $3.922 \times 10^{-8}$ & $-1.152$ \\
        (15,5) & $2.936 \times 10^{-7}$ & $2.977 \times 10^{-9}$ & $-1.522$ \\
        (19,0) & $5.987 \times 10^{-8}$ & $4.606 \times 10^{-8}$ & $-1.334$ \\
        (18,2) & $1.518 \times 10^{-7}$ & $7.226 \times 10^{-9}$ & $-1.668$ \\
        (17,4) & $1.984 \times 10^{-7}$ & $3.889 \times 10^{-11}$ & $-1.991$ \\
        (20,1) & $1.575 \times 10^{-7}$ & $2.985 \times 10^{-7}$ & $-2.180$ \\
        (19,3) & $7.193 \times 10^{-9}$ & $4.232 \times 10^{-9}$ & $-2.543$ \\
        (22,0) & $8.086 \times 10^{-8}$ & $2.691 \times 10^{-11}$ & $-2.786$ \\
        (21,2) & $1.044 \times 10^{-8}$ & $1.153 \times 10^{-9}$ & $-3.137$ \\
        (23,1) & $1.144 \times 10^{-10}$ & $1.105 \times 10^{-9}$ & $-3.800$ \\
        (25,0) & $6.126 \times 10^{-8}$ & $2.903 \times 10^{-11}$ & $-4.522$ \\
    \end{tabular}
    \caption{The interpolating numbers obtained for the RCS and NNSD, together with the corresponding nonlocality detection $\langle \mathcal{B} \rangle < 0$ given the irreps $(p,q)$ that display nonlocality detection with $n=25$ qutrits.}
    \label{tab:rcs_nnsd_n25}
\end{table}

\begin{table}[]
    \centering
    \begin{tabular}{c|c|c|c}
       \textbf{Irrep} & \textbf{RCS} $\boldsymbol{\lambda}$ & \textbf{NNSD (Brody)} $\boldsymbol{\omega}$ & $\langle \mathcal{B} \rangle$ \\
       \hline
    (8, 6) & $4.863\times 10^{-7}$ & $4.110\times 10^{-11}$ & $-0.146$ \\
    (12, 1) & $7.643\times 10^{-8}$ & $9.659\times 10^{-11}$ & $7.548 \times 10^{-10}$ \\
    (11, 3) & $5.245\times 10^{-7}$ & $1.242\times 10^{-09}$ & $-0.146$ \\
    (10, 5) & $2.383\times 10^{-12}$ & $9.416\times 10^{-11}$ & $-0.339$ \\
    (14, 0) & $7.440\times 10^{-8}$ & $1.116\times 10^{-10}$ & $-0.297$ \\
    (13, 2) & $5.358\times 10^{-2}$ & $2.326\times 10^{-01}$ & $-0.405$ \\
    (12, 4) & $1.300\times 10^{-9}$ & $4.165\times 10^{-11}$ & $-0.767$ \\
    (15, 1) & $1.893\times 10^{-6}$ & $3.116\times 10^{-06}$ & $-0.889$ \\
    (14, 3) & $5.978\times 10^{-8}$ & $2.830\times 10^{-09}$ & $-1.178$ \\
    (17, 0) & $4.521\times 10^{-7}$ & $5.325\times 10^{-11}$ & $-1.372$ \\
    (16, 2) & $1.435\times 10^{-12}$ & $5.323\times 10^{-11}$ & $-1.726$ \\
    (18, 1) & $1.667\times 10^{-9}$ & $3.907\times 10^{-08}$ & $-2.319$ \\
    (20, 0) & $7.494\times 10^{-8}$ & $3.916\times 10^{-11}$ & $-3.013$ \\
    \end{tabular}
    \caption{Same as \Cref{tab:rcs_nnsd_n25} but for $n=20$ qutrits.}
    \label{tab:rcs_nnsd_n20}
\end{table}

\subsubsection{Unfolding the energy spectrum}
\label{sec:unfolding}

To compare the nearest-neighbor energy level-space distribution of different operators, it is desirable to normalize it appropriately. The procedure we follow takes the name of spectrum unfolding \cite{Haake2018} and it consists of the following steps.

First, the energy spectrum is sorted so that the energy levels $\{x_i\}_i$ are in ascending order $\{x_1\leq x_2 \leq \ldots \leq x_k\}$. Then, we compute the cumulative distribution function $I_x(E)$ counting the number of energy levels up to energy $E$. This is a discrete function, which it is convenient to interpolate with a continuous polynomial function $\tilde{I}_x(E)$. The goal is now to rescale the sequence $\{x_i\}_i$ into a sequence $\{y_i\}_i$, such that its cumulative function $\tilde{I}_y(E)$ is a straight line. This is achieved by inverting the function $\tilde{I}_x(E)$ and using it to rescale the sequence $\{x_i\}_i$. This spectrum unfolding ensures that the local density of states of the renormalized levels $\{y_i\}_i$ is unity. From the latter sequence we compute the nearest-neighbour energy level spacing as $s_i=y_i-y_{i-1}$, which are used to obtain the NNSD.

\subsection{Looking closer to the anomalous irreps}

In \Cref{fig:RCS_and_NNSD_n25_histograms} we show the RCS histograms for the anomalous irreducible representations that deviate from the otherwise observed Poissonian spacing distribution observed at optimal measurements in the remaining irreps detecting nonlocality. Namely, the cases $(p,q)=(9,8)$ and $(15,2)$ for $n=25$, and $(13,2)$ for $n=20$. For each irrep, we present the RCS distribution obtained both from the measurement settings yielding maximal quantum violation and from randomly generated measurement settings. For the random measurements, we have selected instances whose RCS distributions are very closely fitted by $\lambda \simeq 1$, corresponding to clear Wigner-Dyson (GOE-like) behavior. We emphasize, however, that random measurements do not always yield $\lambda \approx 1$. Rather, they produce a distribution of fitted $\lambda$ values, with $\lambda \simeq 1$ being the dominant case, as shown in \Cref{fig:otherRandProj}. 

Looking at the optimal measurement settings histograms, note that although the fitted RCS parameter is non-zero ($\lambda>0$), the corresponding histograms do not exhibit a pronounced suppression of small spacings. In particular, the first histogram bin does not clearly display level repulsion, in contrast to the random-measurement cases, where level repulsion is clearly visible already at small spacings. This qualitative difference suggests that these spectra lie in a crossover regime between Poisson and Wigner-Dyson statistics, rather than exhibiting fully developed chaotic behavior. This strengthens our interpretation of the residual non-zero $\lambda$ for the optimal measurements as a finite-size effect that might vanish as the irrep dimensions increase towards the asymptotic limit. On the other hand, random measurements generically yield Wigner-Dyson statistics. Hence our conjecture that, in the asymptotic limit, the Bell operators corresponding to measurement settings that maximally violate the Bell inequality are universally integrable within this class of Bell operators.

\begin{figure}
    \centering
    \includegraphics[width=0.46\linewidth]{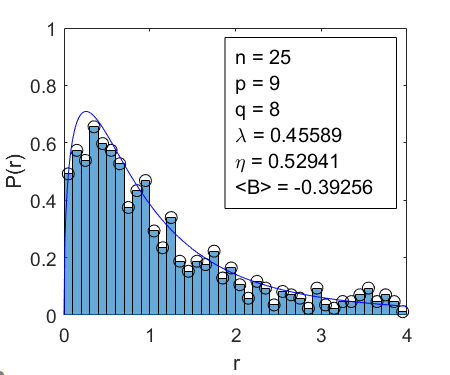}
    \includegraphics[width=0.46\linewidth]{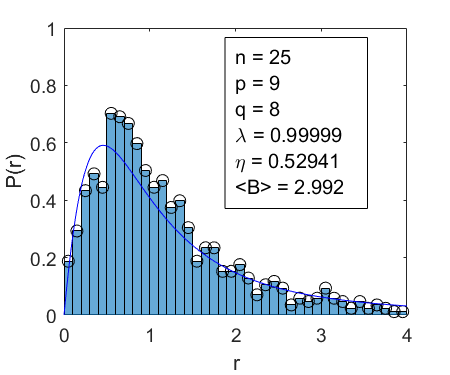}
    \vspace{1em} 
    \includegraphics[width=0.46\linewidth]{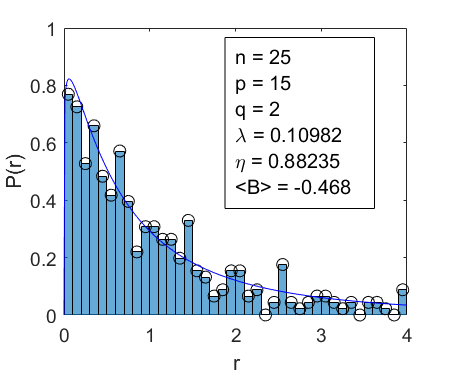}
    \includegraphics[width=0.46\linewidth]{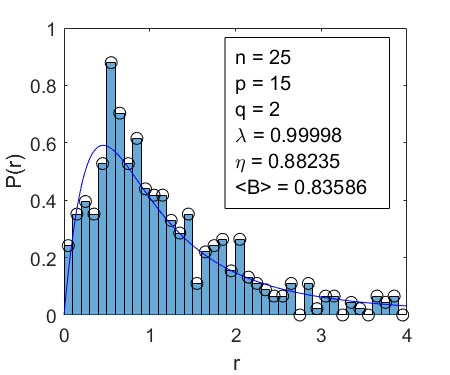}
        \vspace{1em} 
    \includegraphics[width=0.46\linewidth]{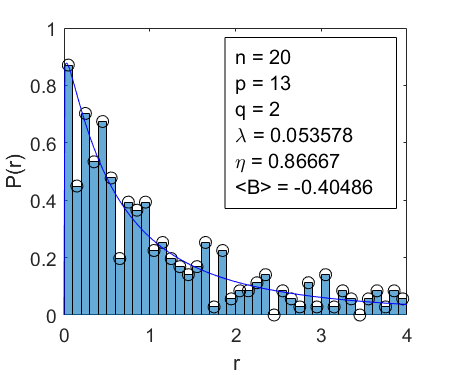}
    \includegraphics[width=0.46\linewidth]{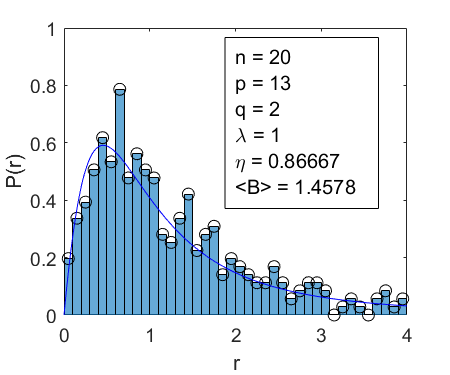}
        
    \caption{RCS histograms for the anomalous irreducible representations $(p,q)=(9,8)$ and $(15,2)$ for $n=25$, and $(13,2)$ for $n=20$. For each irrep, the left panel shows the spectrum obtained using the measurement settings yielding maximal quantum violation, while the right panel shows a representative case obtained from randomly generated measurement settings with an RCS fit $\lambda\simeq 1$. Although the optimal-measurement spectra yield a non-zero fitted parameter $\lambda>0$, they do not exhibit a clear suppression of small spacings (noticeable in the first few bins) and therefore fall into a crossover regime between Poisson and Wigner-Dyson statistics. In contrast, the random-measurement spectra display clear level repulsion at small spacings, characteristic of Wigner-Dyson (GOE-like) behaviour. This observation supports our conjecture that, in the asymptotic limit, the Bell operators corresponding to measurement settings that maximally violate the Bell inequality are integrable within this class of Bell operators.
}
    \label{fig:RCS_and_NNSD_n25_histograms}
\end{figure}

\end{widetext}

\end{document}